\documentclass[a4paper,fleqn,usenatbib]{mnras}

\usepackage[T1]{fontenc}
\usepackage{ae,aecompl}
\usepackage{graphicx}
\usepackage{amsmath}
\usepackage{amssymb}

\title[Velocity gradients]{Non Axisymmetric Relativistic Wind Accretion with Velocity Gradients onto a Rotating Black Hole}

\author[A. Cruz-Osorio and F. D. Lora-Clavijo]{
A. Cruz-Osorio,$^{1}$\thanks{aosorio@astro.unam.mx (ACO)} 
and F. D. Lora-Clavijo$^{2}$\thanks{fadulora@uis.edu.co (FDLC)}
\\
$^{1}$Instituto de Astronom\'{\i}a, Universidad Nacional Aut\'{o}noma de M\'{e}xico, AP 70-264, Distrito Federal 04510, M\'{e}xico.  \\      
$^{2}$Grupo de Investigaci\'on en Relatividad y Gravitaci\'on, Escuela de F\'isica, Universidad Industrial de Santander, A. A. 678,\\
Bucaramanga 680002, Colombia.\\ 
}

\date{Accepted XXX. Received YYY; in original form ZZZ}

\pubyear{2015}

\begin{document}
\label{firstpage}
\pagerange{\pageref{firstpage}--\pageref{lastpage}}
\maketitle

%% ABSTRACT %%
\begin{abstract}

We model, for the first time, the Bondi-Hoyle accretion of a fluid with velocity gradients onto a Kerr black hole, by numerically solving the fully relativistic hydrodynamics equations. Specifically, we consider a supersonic ideal gas, which has velocity gradients perpendicular to the relative motion. We measure the mass and specific angular accretion rates to illustrate whether the fluid presents unstable patterns or not. The initial parameters, we consider in this work, are the velocity gradient $\epsilon_{v}$, the black hole spin $a$, 
the asymptotic Mach number ${\cal M}_{\infty}$ and adiabatic index $\Gamma$. We show that the flow accretion reaches a fairly stationary regime, unlike in the Newtonian case, where significant fluctuations of the mass and angular momentum accretion rates are found. On the other hand, we consider a special case where both density and velocity gradients of the fluid are taken into account. The  spin of the black hole and the asymptotic Newtonian Mach number, for this case, are $a=0.98$ and ${\cal M}_{\infty}=1$, respectively. A kind of flip-flop behavior is found at the early times; nevertheless, the system also reaches a steady state.

\end{abstract}

%% KEYWORDS %%

%% INTRIDUCTION %%
\section{Introduction}

Gas accretion onto black holes is of great interest since this process is associated with the formation and growth of Super Massive Black Holes (SMBHs), X-ray transients and gamma-ray bursts (GRBs).  Observations at distances close to the black holes will be possible due to  the Event Horizon Telescope (EHT) \footnote{http://www.eventhorizontelescope.org}, which is a very long baseline interferometry (VLBI) array consisting of radio telescopes devoted to observe the immediate environment of a black hole with angular resolution comparable to the event horizon. This idea opens a new window on the study of general relativity in the strong field regime \citep{2014ApJ...784....7B}, accretion processes \citep{2015MNRAS.446.1973R} and fundamental black hole physics. In fact, initial EHT observations have made the first detections of structure to scales smaller than the gravitational radius around Sagittarius $A^*$ \citep{2008Natur.455...78D, 2011ApJ...727L..36F}, the SMBH hosting the center of our Galaxy with a mass of $M = 4.1\pm 0.6 \times 10^{6}M\odot$ \citep{2008ApJ...689.1044G}. 

Another project associated with the study of phenomena near to the event horizon of a black hole is GRAVITY \footnote{http://www.mpe.mpg.de/ir/gravity}, which is a new instrument for precision narrow-angle astrometry and interferometric imaging with high sensitivity and accuracy.  This project was designated to explore  physics of the space-time close to the event horizon of the galactic center black hole, for instance, measuring its mass, revealing the details of mass and angular accretion rates, spin and inclination, and also jets. Another interest of GRAVITY is identify if the near-infrared flares from Sagittarius $A^*$ originate from individual hot spots close to the last stable orbit, or from fluctuations in the inner accretion disc, or from a jet \citep{2009NewAR..53..301B, 2010SPIE.7734E..0YG,2011Msngr.143...16E}. Based on these last paragraphs, know about the accretion processes close to the event horizon  is of great significance for the future observations. 

There are many types of black-hole-accretion problems. One of them is the Michel accretion \citep{1972Ap&SS..15..153M}, which is the relativistic extension to the Bondi accretion problem \citep{1952MNRAS.112..195B}, consisting on the radial accretion of a homogeneously distributed gas surrounding a spherically symmetric black hole, under the assumption the space-time is fixed, which means the fluid is not sufficiently massive to change the black hole mass. This solution depends on two free parameters, the density and pressure, at a point where the fluid velocity equals the sound speed. A recent work has reformulated this accretion problem as a dynamical one, and under assumptions on the fluid equation of state, the authors determine the local and global qualitative behavior of its phase flow, proving that for any given positive particle density number at infinity, there exists a unique radial steady-state accretion flow, which is regular at the horizon \citep{2015CQGra..32o5006C}. On the other hand, the authors also study the spherical and non-spherical linear acoustic perturbations of the Michel flow \citep{2015PhRvD..91j4012C} as well as the properties of a polytropic fluid, which is radially accreted into a Schwarzschild black hole \citep{2015arXiv151107728C}. In \cite{2011MNRAS.415..225G,2011MNRAS.416.3083G} the conditions for stable radial accretion and runaway accretion of an ideal gas have been studied. More recently, radial accretion in the non-linear regime has been used to study primordial black holes during the radiation-dominated era, showing that they can grow up to $10^3-10^6 M\odot$ \citep{2013JCAP...12..015L}. In the same regime, \cite{2015ApJS..218...24L} studied the accretion of self-interacting dark matter into a supermassive black hole for the particular case of radial flows.
 
Another type of accretion problem is known as the Bondi-Hoyle-Lyttleton (BHL) accretion \citep{1939PCPS...35..405H,1944MNRAS.104..273B}, which consists of a homogeneous fluid moving toward a central compact object. The morphology of this process depends on whether the gas velocity is supersonic or not, that is, a shock cone appears when the fluid is supersonic and in the other case not. This process shows interesting properties when considered within the Newtonian and relativistic regimes, which have been explored based on several numerical studies. A summary of results, under Newtonian gravity, can be found in \citep{2004NewAR..48..843E,2005A&A...435..397F}. The first works in the relativistic regime, using axial and equatorial symmetries, were performed by \cite{1989ApJ...336..313P,1998MNRAS.298..835F,1998ApJ...494..297F,1998ApJ...507L..67F,1999MNRAS.305..920F}, where different patterns close to the black hole were studied using numerical evolutions. On the other hand, relativistic BHL has been used in the astrophysical context, for instance, in \citep{2011MNRAS.412.1659D,2013MNRAS.429.3144L} the authors found vibrations in the shock cone, which may potentially explain high energy process like QPOs. \cite{2011MNRAS.412.1659D, 2012MNRAS.426.1533D} showed a flip-flop behavior of the gas surrounding a rotating black hole. Thereafter, in \cite{2012MNRAS.426..732C} was shown that when Boyer-Linquist coordinates are used, the shock cone oscillates in a flip-flop fashion,whereas when using Kerr-Schild coordinates such behavior does not appear. More realistic scenarios introduce astrophysically relevant ingredients like magnetic fields \citep{2011MNRAS.414.1467P}, radiative terms \citep{2011MNRAS.417.2899Z} and inclusive ultra-relativistic processes associated with the growing of primordial black holes during the radiation era  \citep{2013MNRAS.428.2171P,2013AIPC.1548..323C}.Recently, \citep{2015ApJS..219...30L} showed a numerical study of the Bondi-Hoyle accretion with density gradients in the fully relativistic regime. In this work, the authors presented different astrophysical scenarios, associated with stellar, intermediate-mass and supermassive black holes, where Bondi-Hoyle with density gradients can take place. 

In this manuscript, we present, for the first time, a numerical study of the relativistic Bondi-Hoyle accretion with velocity gradients around a rotating black hole. Specifically, we show the gas morphology and some of its properties. This work is a complement to the work \citep{2015ApJS..219...30L}, where a non-homogeneous fluid was considered to study the Bondi-Hoyle accretion in the fully relativistic regime. According to \cite{1997A&A...317..793R}, in the Newtonian regime, the models exhibit active unstable phases and the mass and specific angular momentum accretion rates fluctuate with time. Therefore, in order to confirm or discard these fluctuations in the relativistic regime, the space of parameters we explore is:  velocity gradient parameter $(\epsilon_{v}=0.1,~0.2,~0.3)$, spin of the black hole $(a=0,~0.5,~0.98)$, asymptotical Mach number $({\cal M}_{\infty}=1,~3,~4,~5)$ and adiabatic index $(\Gamma = 4/3,~5/3,~2)$. In all the cases, we found a dragging on the shock cone due to the velocity gradient. On the other hand, all the considered models reach a fairly stationary flow pattern, as in the case of density gradients \citep{2015ApJS..219...30L}. Finally, we show a special case, where both density $(\epsilon_{v}=0.3)$ and velocity $(\epsilon_{\rho}=0.5)$ gradients are assumed for a black hole with a very high spin $(a=0.98)$. In  this case, we found a kind of flip-flop behavior at the early times; however, the system also reaches a fairly steady state.

The plan of the paper is as follows. Section \ref{sec:ENS} briefly describes the relativistic hydrodynamic equations as well as the numerical methods employed to carried out the simulations. Section \ref{sec:IData} describes the initial data corresponding to a fluid with density and velocity gradients. Section \ref{sec:NResults} shows our numerical results. Finally, in Section \ref{sec:conclus}, we summarize our main results and discuss them.  We use units in which $G=c=1$ in such way that length and time are measured in units of $M$.

%% EQUATIONS AND NUMERICAL SETUP%%
\section{Equations and Numerical Setup }
\label{sec:ENS}

\subsection{Relativistic Hydrodynamic Equations}

The relativistic hydrodynamic equations, in a curved space-time, can be derived from the local conservation of the energy-momentum tensor $\nabla_{b}T^{a b}=0$ and the conservation of rest mass $\nabla_{a} (\rho u^{a})=0$, where $\rho$ is the rest  mass density, $u^{a}$ are the components of the 4-velocity of the fluid elements and $\nabla_a$ stands by the covariant derivative associated with the 4-metric $g^{ab}$. We use a perfect fluid to model the relativistic gas, with energy-momentum tensor $T^{ab} = \rho h u^a u^b + p g^{a b}$, where $p$ is the thermal pressure and $h = 1 + p \Gamma / (\Gamma -1)\rho$ is the specific enthalpy. We have assumed the gas obeys an ideal equation of state $p = (\Gamma - 1)\rho e$, where $e$ is the specific internal energy density and $\Gamma$ the adiabatic index. In order to see in more detail the equations used in this work, we refer to the article \cite{2015ApJS..219...30L}. Latin index run from 0 to 3.

These equations are solved onto a rotating black hole in Kerr-Schild coordinates, which allow that the fluid naturally falls into the event horizon \cite{2012MNRAS.426..732C}.  On the other hand, we will restrict the polar coordinate to $\theta = \pi/2$, thereby restricting the flow to the equatorial plane of the black hole. The axis of rotation of the rotating black hole is orthogonal to the direction of the fluid flowing from infinity. The infinitely thin-disk approximation is highly restrictive and is used to gain a basic understanding of the flow morphology when velocity gradients are considered. To obtain a full understanding of the nature of accretion onto a rotating black hole, a full three-dimensional evolution will be necessary. Finally, we compute the mass and angular accretion rates in a spherical detector located close to the event horizon, see \cite{2015ApJS..219...30L}.

\subsection{Numerical Setup}

This work is carried out using the CAFE magnetohydrodynamics code \cite{2015ApJS..218...24L}. CAFE includes different schemes for solving the relativistic hydrodynamics and magnetohydrodynamics equations as well as the Newtonian magnetohydrodynamics \cite{2015arXiv150900225G}. In particular, the code is based on high-resolution shock-capturing schemes, uses the HLLE flux formula  combined with minmod, mc and weno5 reconstructors and in order to preserve the magnetic field divergence  it uses or Flux Constrained Transport  or Divergence Cleaning methods. Even though the relativistic magnetohydrodynamics in CAFE is written on the Minkowski space-time, the relativistic hydrodynamics routine solves the equations on a curved space-time in equatorial and axial symmetries \cite{2012MNRAS.426..732C, 2013MNRAS.429.3144L}. In order to update in time, CAFE uses the method of lines combined with a third order TVD (Total Variation diminishing) or fourth order Runge-Kutta time integrators. Finally, in order to get the primitive variables in terms of the conservative ones a hybrid bisection Newton-Raphson algorithm  is used in each time step within the evolution scheme. 

In this particular work, we use a fourth order Runge-Kutta time integrator combined with the HLLE approximate Riemann solver and the minmod linear reconstructor. The numerical simulations are carried out on the numerical domain  $[r_{min},r_{max}] \times [0,2\pi)$, where the grid is uniformly divided with the cell size $(\Delta r, \Delta \phi)$ =$(0.117,0.024)$. We choose the interior boundary $r_{min}$ to be inside the black hole horizon, where we apply an excision. The exterior boundary $r_{max}$ is divided into two halves, one in which the gas enters the domain where we apply inflow boundary conditions, and a second half where the gas leaves the domain and where we apply outflow boundary conditions there. The numerical grid is uniformly spaced with resolutions $\Delta r, \Delta \phi$. We use a constant time step given by $\Delta t = C~min(\Delta r,\Delta\phi)$, where $C$ is the Courant-Friedrich-L\"{o}wy condition, which in our case is $C = 0.2$. 

%% INITIAL DATA %%
\section{Initial Data}
\label{sec:IData}

The initial data corresponds to a non homogeneously distributed gas which fills whole the domain and moves on the equatorial plane along the $x$ direction with velocity gradients. The initial velocity profile $v^i = (v^r,v^{\phi},0)$, in polar coordinates, is given by the expressions 
\begin{align}
\nonumber v^{r}&= v_{ini}\left(H_{1} \cos \phi+H_{2} \sin \phi \right), \\\nonumber 
v^{\phi}&= v_{ini} \left (H_{4} \cos \phi - H_{3} \sin \phi \right),
\end{align}
where $H_i$ ($i = 1, 2, 3, 4$) are functions associated with the components of the space-time metric; for more detail of this see  \citep{2012MNRAS.426..732C,2015ApJS..219...30L}. Now, following Rufferts's  work \citep{1995A&A...295..108R,1997A&A...317..793R,1999A&A...346..861R}, we chose $v_{ini}$ and the initial density profile $\rho_{ini}$ as 
\begin{align}
v_{ini} &= v_{\infty} \Big\{ 1- \frac{1}{2}\tanh\Big[2\epsilon_{v}\frac{(r \sin[\phi] + a \cos[\phi])}{r_{a}}\Big] \Big\}, \\\rho_{ini} &= \rho_{0} \Big\{ 1- \frac{1}{2}\tanh\Big[2\epsilon_{\rho}\frac{(r \sin[\phi] + a \cos[\phi])}{r_{a}}\Big] \Big\}, 
\end{align}
where $\rho_{0}$ is a constant density value, $v_{\infty}$ is an asymptotic initial velocity, $a=J/M$ is the black hole rotation parameter, $\epsilon_{\rho}$ and $\epsilon_{v}$ are the parameter specifying the magnitude of the density and velocity gradients respectively and $r_a = M / \left( v_{\infty}^2 + c_{s \infty}^2 \right)$ is the accretion radius, which is defined in terms of the asymptotic values of the sound speed $c_{s_\infty}$  \cite{1989ApJ...336..313P}. The initial data, we consider in this work, can be summarized in Table \ref{tab:params}, where the relativistic speed of sound is set to $c_{s \infty}= 0.1$.

\begin{table}
	\centering
	\caption{In this table, we summarize the parameters used, as initial configurations, in our numerical simulations. In all the calculations, we fix the relativistic sound speed to $c_{s \infty}= 0.1$.}
	\label{tab:params} 
 \begin{tabular}{|c|cc|cc|cc|c|c|}\hline\hline
 \multicolumn{1}{c|}{Model} & \multicolumn{2}{c|}{a=0} & \multicolumn{2}{c|}{a=0.5} & \multicolumn{2}{c|}{a=0.98} & \multicolumn{1}{c|}{${\cal M}_{\infty}$}& \multicolumn{1}{c|}{${\Gamma}$} \\ 
              &  $\epsilon_{v}$ & $\epsilon_{\rho}$ & $\epsilon_{v}$ & $\epsilon_{\rho}$ & $\epsilon_{v}$ & $\epsilon_{\rho}$  \\ \hline 
 $M_1~~$  &  $0.1$ &  0.0  &  -  &  -  & - & -   &  4   &5/3 \\ 
 $M_2a$    &   -        &    -   & $0.1$ &  0.0  & - & - &   3 &5/3 \\ 
 $M_2b$    &   -        &  -     & $0.1$ &  0.0  & - & - &   4 &5/3 \\ 
 $M_2c$    &   -        &  -     & $0.1$ &  0.0  & - & -  &  5 &5/3 \\ 
 $M_3~~$ &  $0.2$  &  0.0 &  - &  -  & - & - &4&5/3 \\ 
 $M_4a$    &   -        &  -    & $0.2$ &  0.0  & - & - &3&5/3\\ 
 $M_4b$   &   -         &  -    & $0.2$ &  0.0  & - & - &4&5/3\\ 
 $M_4c$   &   -         &  -     & $0.2$ &  0.0  & - & - &5 &5/3\\ 
 $M_5~~$ &  $ 0.3 $&  0.0  & - & -  & - & -  &4   &5/3  \\ 
 $M_6a$   &  -         &  -     & $0.3$ &  0.0  & - & -  &3  &5/3 \\ 
 $M_6b$   &  -         &  -     & $0.3$ &  0.0  & - & -  & 4  &5/3   \\ 
 $M_6c$   &  -         &  -     & $0.3$ &  0.0  & - & -  & 5   &5/3 \\ 
 $M_7~~$ &    -       &   -    &  -  &  -  & 0.3 & 0.5  & 1  &5/3 \\
 $M_8~~$ &    -       &   -    &  0.2 &  0.0  & - & -  & 4  &4/3 \\ 
 $M_9~~$ &    -       &   -    &  0.2 &  0.0 & - & -  & 4  &2 \\ \hline \hline
\end{tabular}  
\end{table}

%% NUMERICAL RESULTS%%
\section{Numerical Results}
\label{sec:NResults}

\begin{figure*}
\includegraphics[width=9cm,height=9cm]{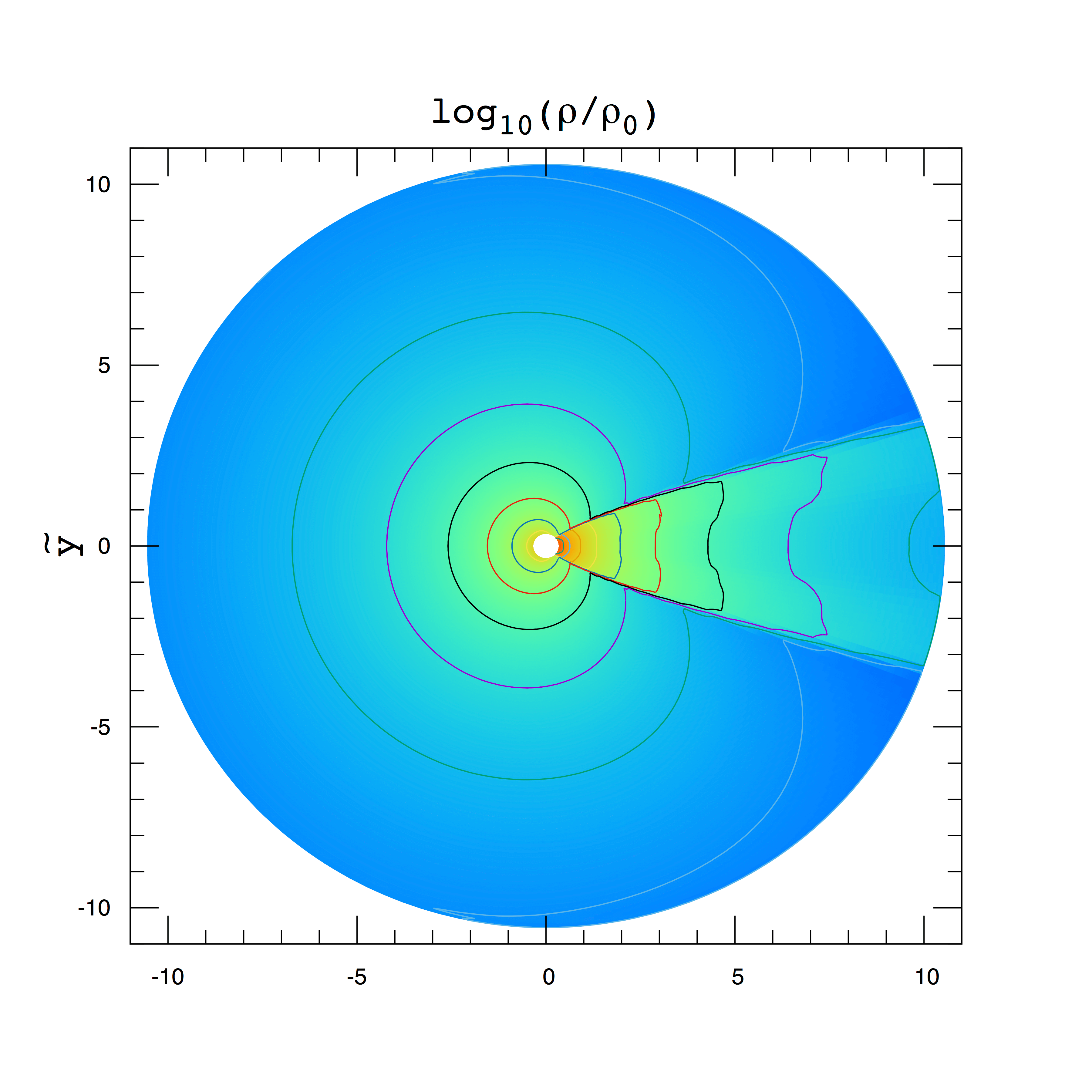} \hspace{-1.2cm} \includegraphics[width=9cm,height=9cm]{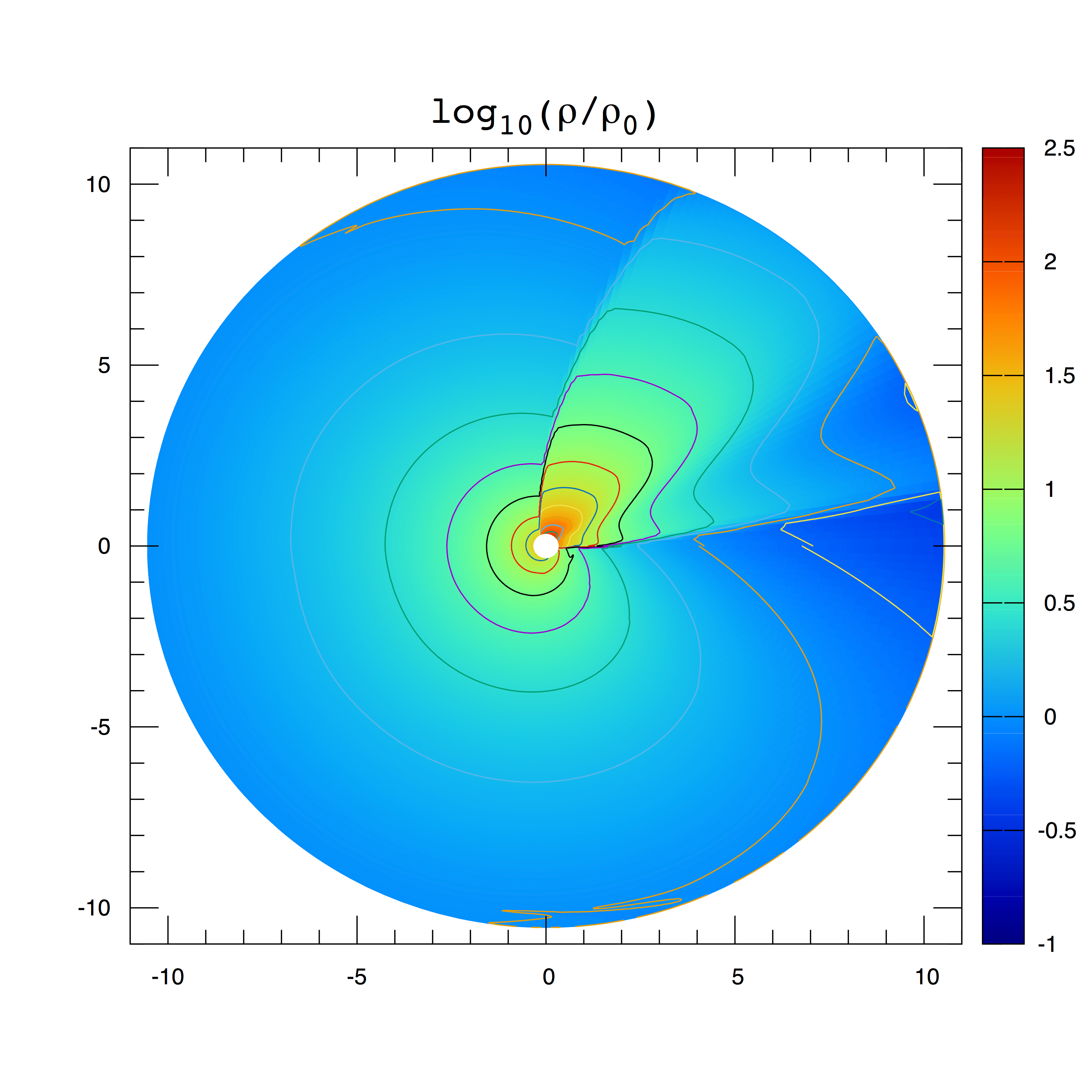}\\
\vspace{-1.25cm}
\noindent \includegraphics[width=9cm,height=9cm]{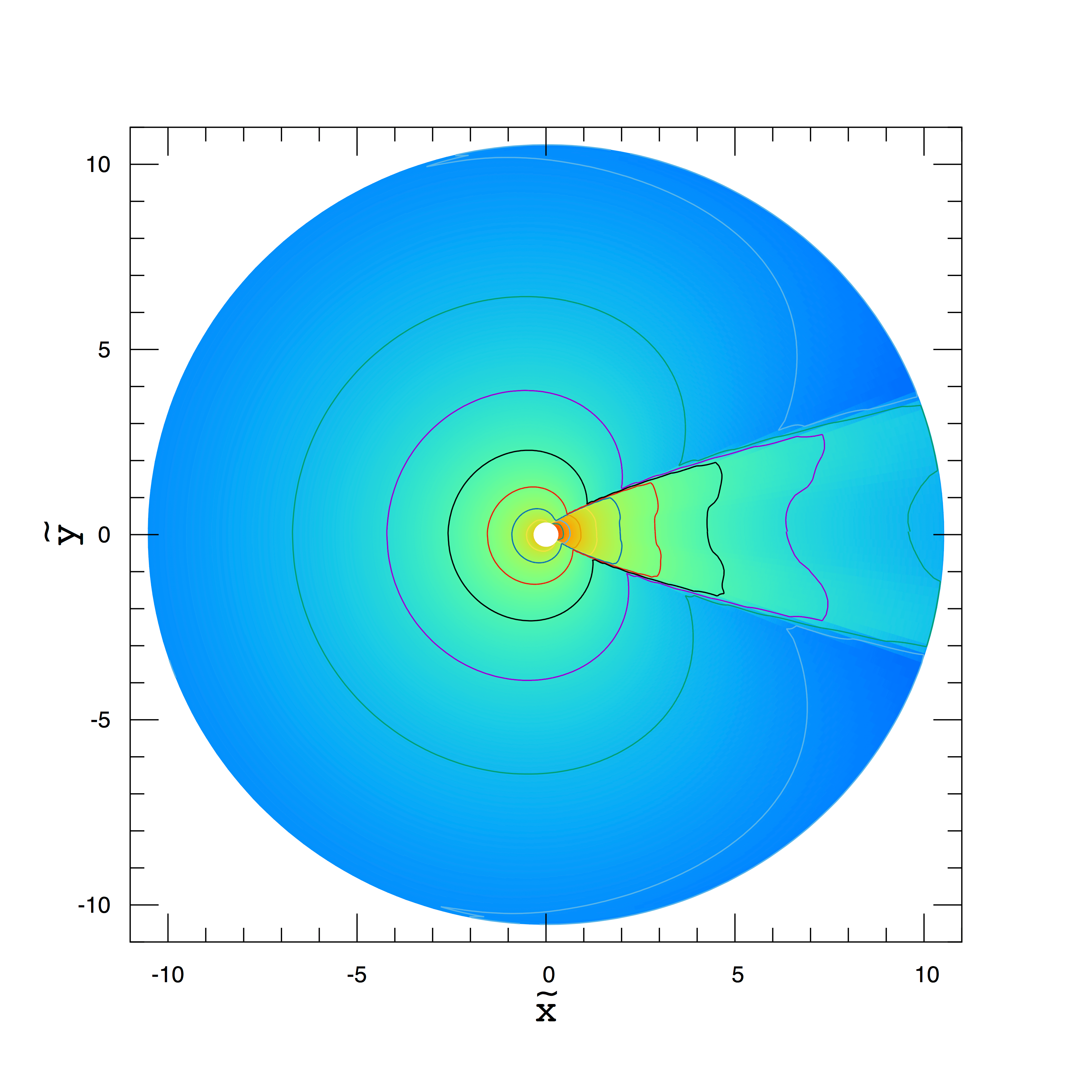}\hspace{-1.05cm}\includegraphics[width=9cm,height=9cm]{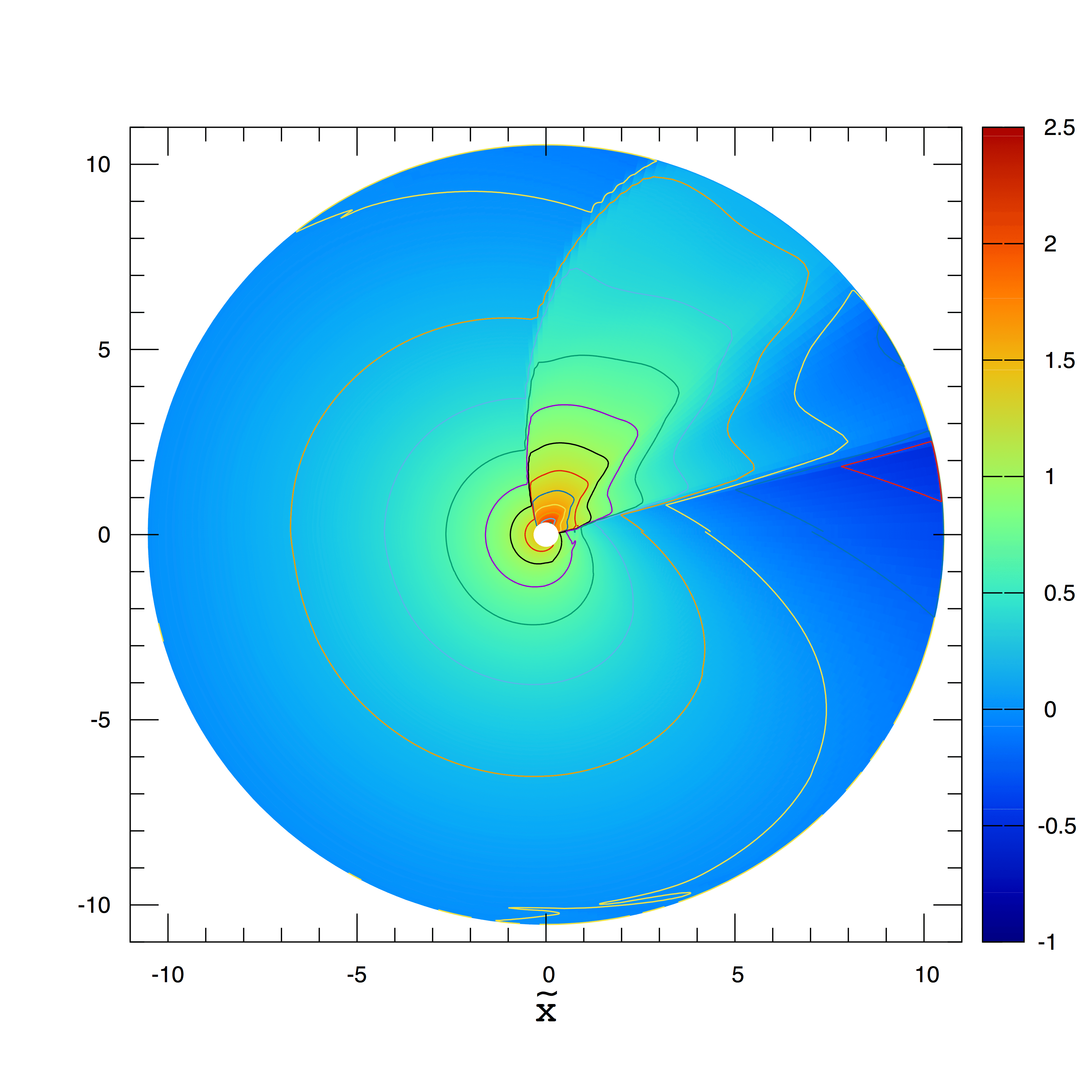}
\caption{ Logarithm of the rest mass density in geometric units at $t=2000M$, once the system reaches a stationary regime. The models correspond to a fluid moving with Newtonian Mach number  ${\cal M}_{\infty} = 4$. The left column shows results, with non-velocity gradient, for two values of the spin of the black hole, $a=0$ (top) and $a=0.5$ (bottom), while the right column presents the results obtained with a velocity gradients parameter of $\epsilon_{v} = 0.3$, for the same spins. It can be appreciated a dragging of the shock cone due to the velocity gradients, fact that is more notorious when the velocity gradient is increased. The coordinates are normalized by the accretion radius as $\tilde{x}= x/r_{a}$ and  $\tilde{y}= y/r_{a}$. }
    \label{fig:2}
\end{figure*}

In this section, we present a numerical study of the effects in the accretion of a relativistic wind with velocity gradients onto a rotating black hole, which is described by the Kerr line element in Kerr-Schild coordinates. In particular, we have studied the mass and angular momentum rates for different parameters. Specifically, we have focused in the spin of the black hole $a$, and the velocity-gradient parameter of the gas, $\epsilon_{v}$.  In order to illustrate the morphology,  we show a comparison between the case a with non-velocity gradient ($\epsilon_{v}=0$ ) and the one with $\epsilon_{v}=0.3$ (Figure \ref{fig:2}), once the system has reached a steady state. The mass and angular momentum rates are shown in Figures \ref{fig:4} and \ref{fig:5}, respectively. As we can see from these models a nearly stationary pattern is reached. Finally, we considered a special case, where not only velocity gradients are considered but also strong density gradients $\epsilon_{\rho}$, see Figures \ref{fig:3} and \ref{fig:6}. 

\subsection{Morphology}

In this section, we show, for the first time, the morphology of a relativistic supersonic wind, which is moving toward a rotating black hole with velocity gradients. Despite this first work is not carried out in full 3D, slab symmetry gives us a basic understanding of the morphology when velocity gradients are considered. This first intent is the first one toward more realistic scenarios (more realistic 3D simulations). In order to illustrate the morphology of the supersonic relativistic Bondi-Hoyle accretion with velocity gradients, we compare in Figure \ref{fig:2} the results obtained with and without gradients for different spins of the black hole. Specifically, in the left column of this figure, we show the case with a uniform velocity gradient with spin parameters $a=0$ (top) and $a=0.5$ (bottom), while in the right column, we present, for the same spin parameters respectively, the case with $\epsilon_{v} = 0.3$. The snapshots represent the logarithm of the gas density in geometrical units  at the time $t=2000M$, once a steady state is achieved.

The most notorious difference on the morphology, between the system with uniform velocity profile and the one with gradients, is the dragging produced on the shock cone when $\epsilon_{v} \neq$ 0 is considered; such effects are more significant when the velocity gradient is increased. Of all the models considered, the model $M_6$ has greater dragging than the others, showing a bigger opening angle of the shock cone. On the other hand, the spin effects can be appreciated very close to the black hole while the gradients have a high significance in all numerical domain.

\begin{figure}
\includegraphics[width=8.5cm,height=6.5cm]{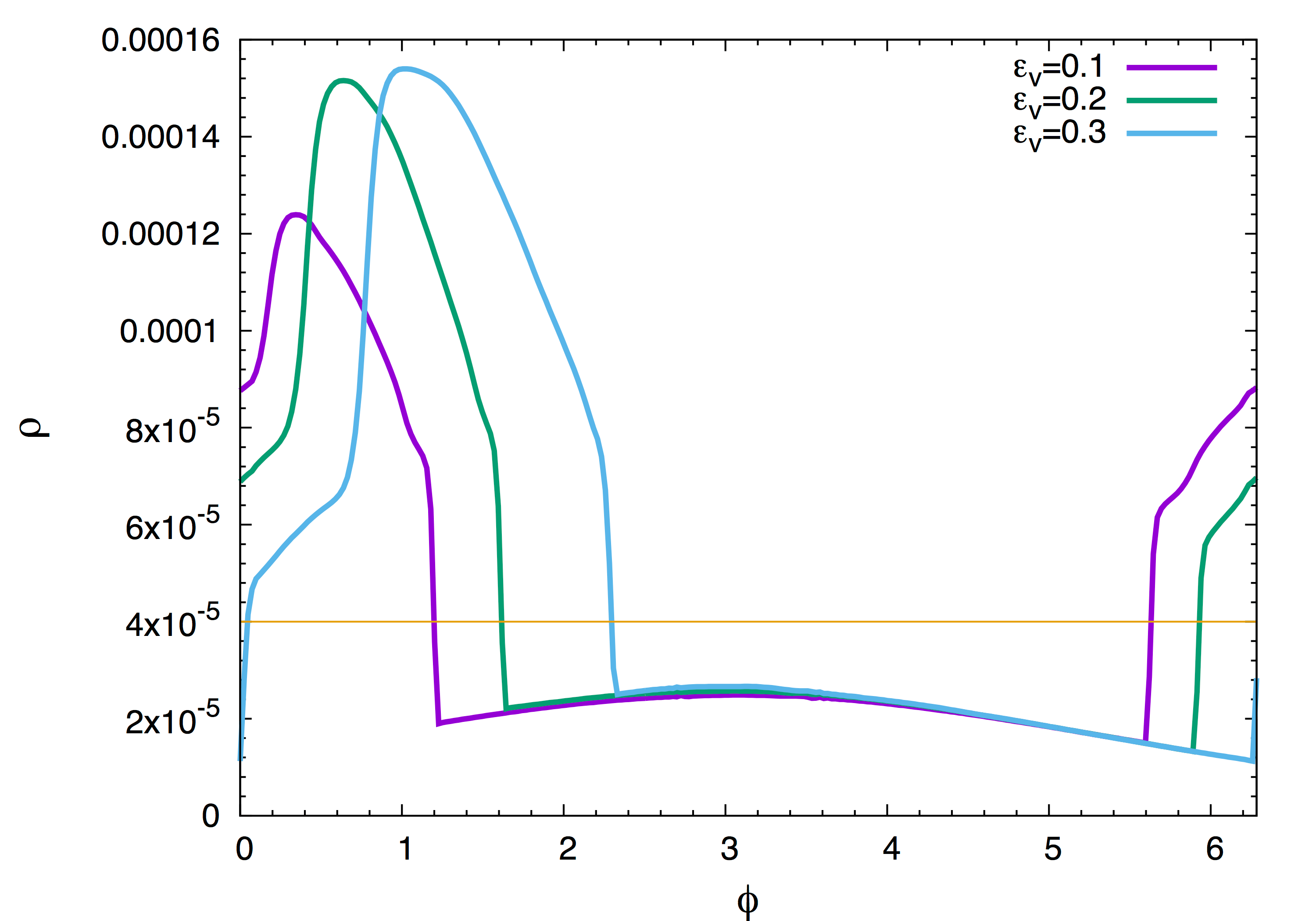} 
    \caption{Rest mass density profile showing the location of the shock cone for different values of the velocity gradient parameter $\epsilon_{v}$. The asymptotic flow velocity, spin of the black hole and adiabatic index are $v_{\infty}=0.4$, $a=0.5$ and $\Gamma = 5/3$, respectively. As we can observe from this figure, the bigger the velocity gradient parameter the bigger the dragging and the opening angle of the shock cone. The plot  is presented once the system has reached a stationary flow pattern.}
    \label{fig:10}
\end{figure}

 In order to illustrate the different effects over the shock cone due to the velocity gradients, we show in 
Figure \ref{fig:10}, once the system has reached a steady state, the rest mas density profile as a function of the angle $\phi$, measured by a  detector located close to the event horizon. We compare the dragging effect and the opening angle of the shock cone for three different values of $\epsilon_{v}=0.1, 0.2 0.3$, that is for the models $M2b,~M4b,~M6b$. As we can see from this figure, when the velocity gradient parameter increases the dragging effect is more noticeable and the opening angle is bigger. Specifically, for $\epsilon_{v}=0.1$, $\epsilon_{v}=0.2$ and $\epsilon_{v}=0.3$ the opening angles are  $~1.8437$, $~1.96148$ and $~2.2508$, respectively. The same behavior is observed for the models $M2a,~M2c,~M4a,~M4c,~M6a,~M6c$.

The morphology of the rest mass density gives us a clear vestige about the effects of the gradients in the accretion process. Such gradients are able to change the behavior of the wind along the evolution giving a general evidence that the accretion of  winds can be totally different due to the non-uniform properties of the  gas in the interstellar medium. 

\subsection{Accretion rates}

In this subsection, we focus on the analysis of our numerical results obtained for the mass and angular momentum accretion rates, which are computed following the expression presented in our work \cite{2015ApJS..219...30L}. These quantities are measured  in a detector located close to the event horizon of the black hole. 

\begin{figure}
\includegraphics[width=\columnwidth]{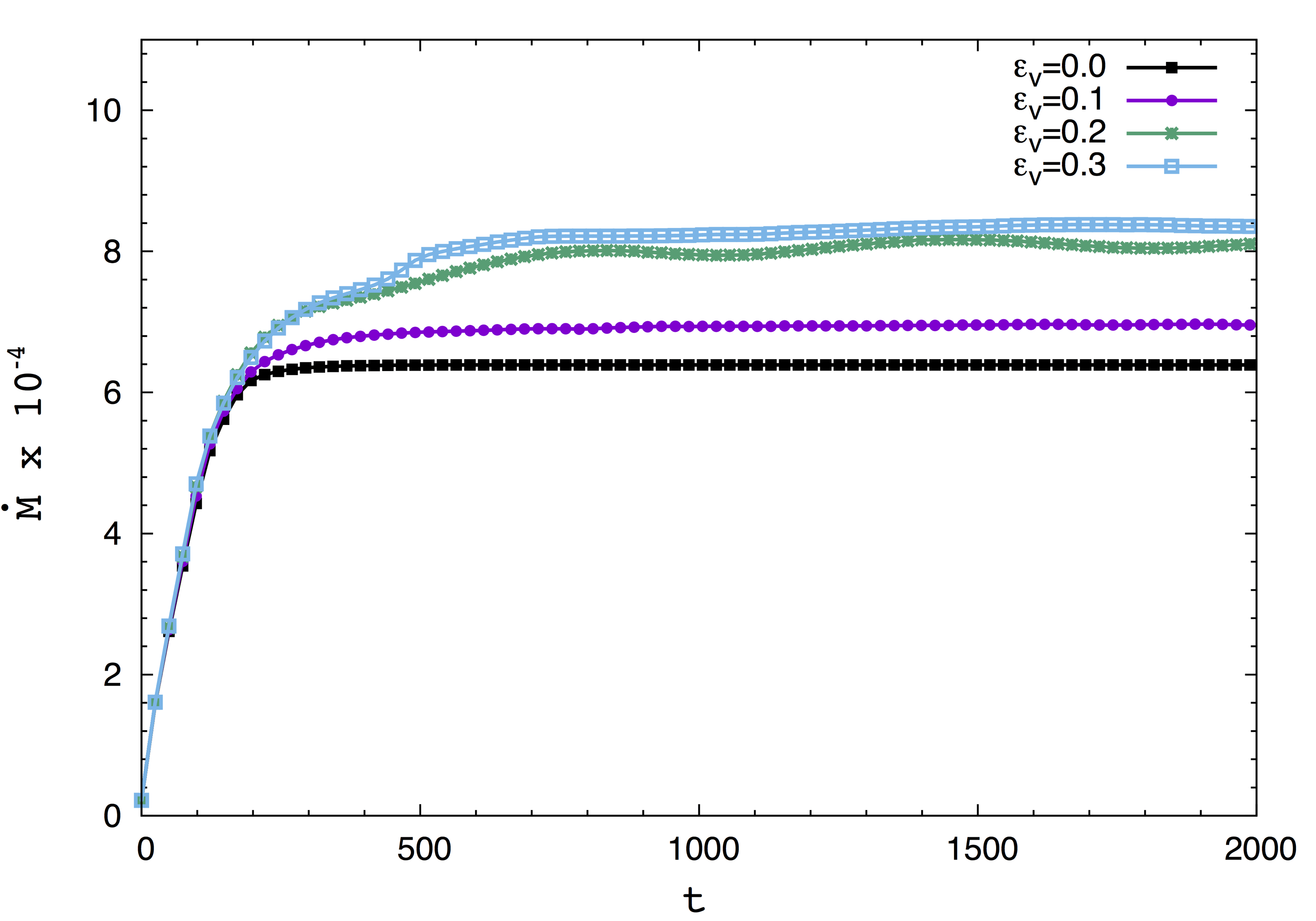}\\ 
\includegraphics[width=\columnwidth]{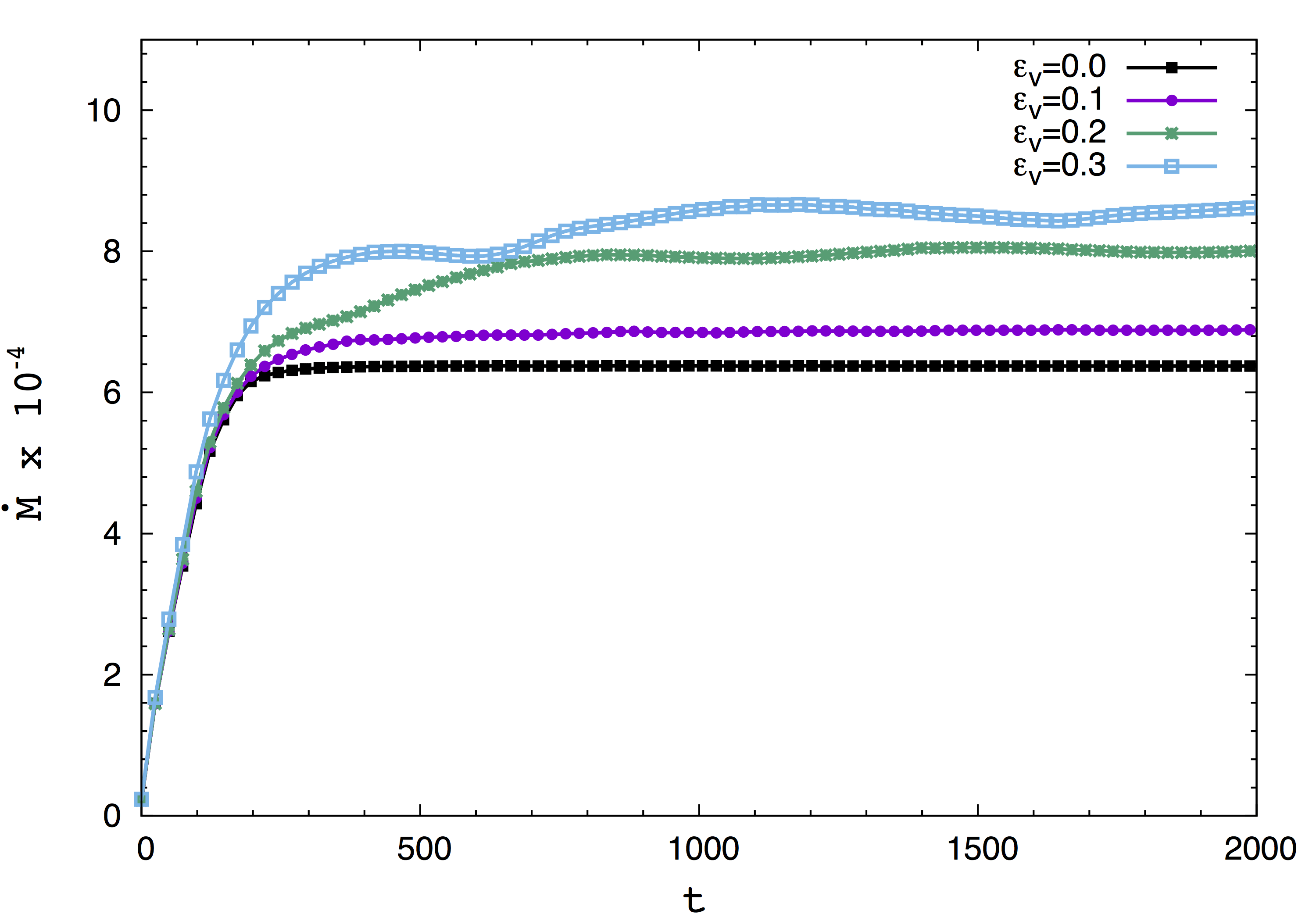}
    \caption{Evolution of the mass accretion rates, measured by a detector located close to the event horizon, for different values of the  gradient velocity parameter $\epsilon_{v}$. The upper panel shows $\dot{M}$ for a Schwarzschild black hole, while the lower one corresponds to a rotating Kerr black hole with spin parameter $a=0.5$. In each figure, we show four different values of the velocity gradient parameter $\epsilon_{v} = 0, 0.1, 0.2, 0.3$, corresponding to the models $M_{1}-M_{6}$ in the Table  \ref{tab:params}, respectively. Even thought in the models  $M_5$ and $M_6$, $\dot{M}$ is a bit unstable than the others, it is possible to set that the bigger the velocity gradient is the bigger the mass accretion rate.}
    \label{fig:4}
\end{figure}

The mass accretion rates are presented in Figure \ref{fig:4}, where we can observe that the behavior along the time evolution reaches rather quickly to a steady state for all the models; regardless of the different values of $\epsilon_{v}$  that are considered. These values range from $10 \%$ to $30 \%$ over one accretion radius or gradient values $\epsilon_{v}=0, ~0.1,~0.2$ and $0.3$. These results differ from the Newtonian case \cite{1997A&A...317..793R}, where an unstable and decreasing mass accretion rates were encountered. We show in the top panel of this figure, the case corresponding to a gas moving towards a Schwarzschild black hole, while in the bottom one, the case corresponding to a gas moving toward a Kerr black hole with spin parameter $a=0.5$. We found that the spin of the black hole does not affect too much the mass accretion rates, unlike the gradient velocity parameter, which produces an increment in the mass accretion rate. Even though the mass accretion rates for the cases $\epsilon_{v}=0.2$ and $\epsilon_{v}=0.3$ are similar, we observe the value of $\dot{M}$ is bigger for $\epsilon_{v}=0.3$. Thereby, we can set that the greater the value of $\epsilon_{v}$ is the greater the mass accretion rate. However, it is worth mentioning that when $\epsilon_{v}$ increases the system becomes a bit more unstable. Even so, all the models considered reach a fairly steady state, like in the case of non-velocity gradients  \citep{1998MNRAS.298..835F,1998ApJ...494..297F,1998ApJ...507L..67F,1999MNRAS.305..920F}.

\begin{figure}
\includegraphics[width=8.5cm,height=6.5cm]{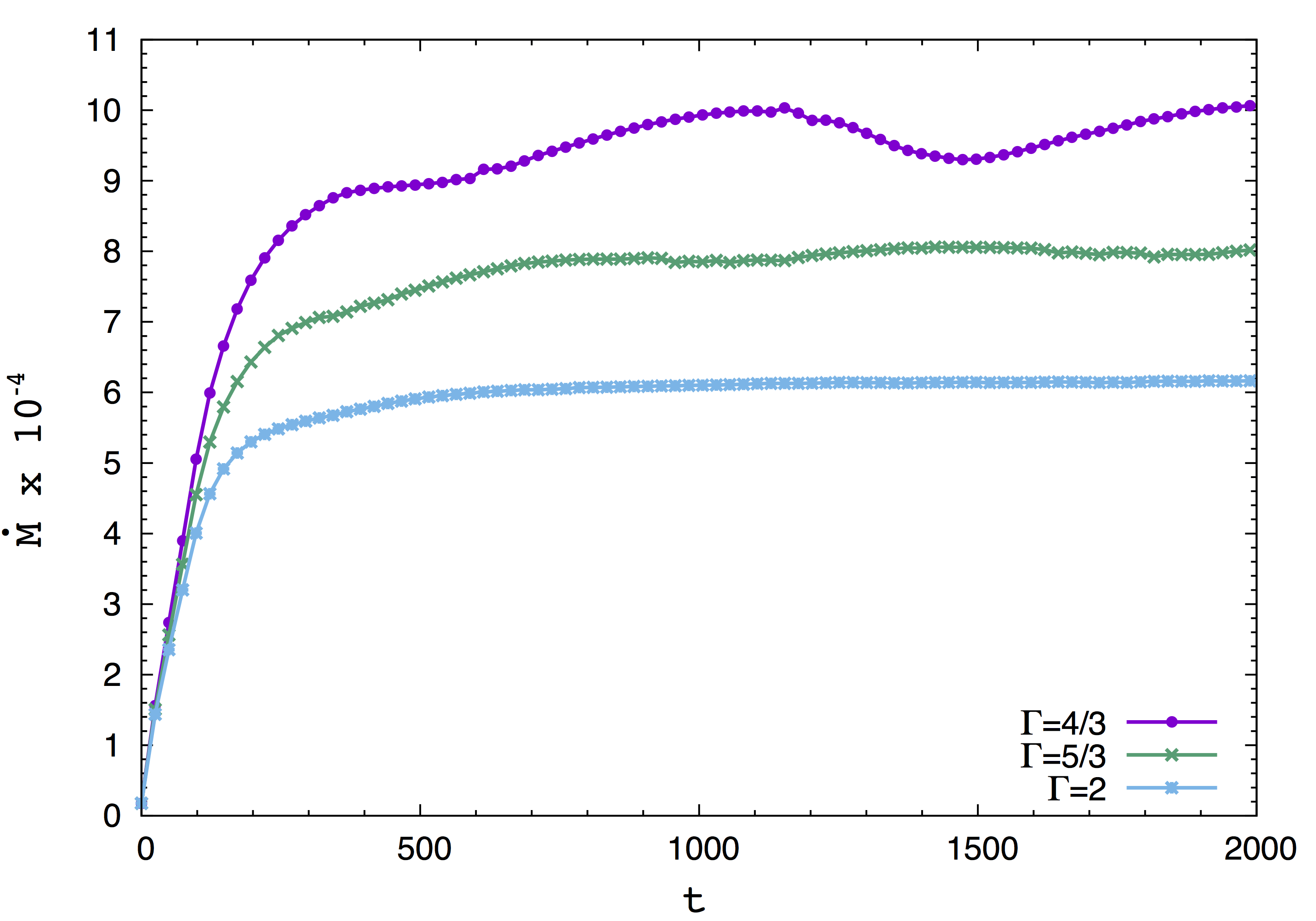} 
    \caption{ Mass accretion rates for a relativistic wind (${\cal M}_{\infty} = 4$) with velocity gradient parameter 
    $\epsilon_{v} = 0.2$ moving towards a rotating black hole (a=0.5) with different values of the adiabatic index. When the adiabatic index increases the mass accretion rate decreases, that is the system reaches faster to a steady state for bigger values of $\Gamma$.}
    \label{fig:9}
\end{figure}

On the other hand, in order to illustrate the effects for different values of the asymptotical Mach number and the adiabatic index, we show the figures \ref{fig:7} and \ref{fig:9}, respectively.  In both cases the spin of the black holes is $a=0.5$. Specifically, in Figure  \ref{fig:7}, we present the mass accretion rates (left column) for a relativistic fluid  with Mach number ${\cal M}_{\infty}=3$ (left top) and ${\cal M}_{\infty}=5$ (left bottom). As we can see from both cases, the relativistic flows also reach a stationary pattern and like in the case of ${\cal M}_{\infty}=4$ we can state that the bigger the velocity gradient parameter the bigger the mass accretion rate. Now, for a given value of $\epsilon_{v}$, it is worth mentioning from Figure \ref{fig:4} (bottom) and Figure \ref{fig:7} (left column) that the mass accretion rate increases when the mach number decreases. Finally, we also show in Figure \ref{fig:9} a comparison of $\dot{M}$ for different values of $\Gamma$ for a relativistic fluid with velocity gradient $\epsilon_{v}=0.2$. In this case, as it was expected, for bigger values of the adiabatic index the system reaches faster a steady state. 

Because the space-time considered in this paper corresponds to a rotating black hole, we are also interested to know the effects of the velocity gradient on the angular momentum accretion rates. Specifically, we compare in Figure \ref{fig:5}, the behavior of the angular momentum accretion rate, for a black hole with non-spin ($a=0$) (top panel) and for one with spin ($a=0.5$) (bottom panel). We reproduce the case where non-velocity gradients are considered \cite{1999MNRAS.305..920F}, that is, the angular momentum accretion rate is zero in the case with spin zero (top panel - dot black line) and  $\sim -1.6$ in the case with spin $a=0.5$ (bottom panel - dot black line). In the non-uniform case, we can see the fluctuations are more notorious in the cases $\epsilon_{v}=0.2$ and $\epsilon_{v}=0.3$, however as we mentioned before the system reaches a stationary flow pattern.  We also see that the bigger the velocity gradient parameter is the bigger angular momentum accretion rate. Finally in Figure  \ref{fig:7} (right column), we show the momentum accretion rate for different values of the asymptotical Mach number, ${\cal M}_{\infty}=3$ (right top) and ${\cal M}_{\infty}=5$ (right bottom). As we can see, again the oscillations are more noticeable for the cases $\epsilon_{v}=0.2$ and $\epsilon_{v}=0.3$,  even so the behavior is similar as in the case ${\cal M}_{\infty}=4$, that is, the system reaches to a fairly stationary flow pattern.
 
\subsection{Special case $M_7$}

\begin{figure*}
\vspace{-0.5cm}
\includegraphics[width=0.75\columnwidth]{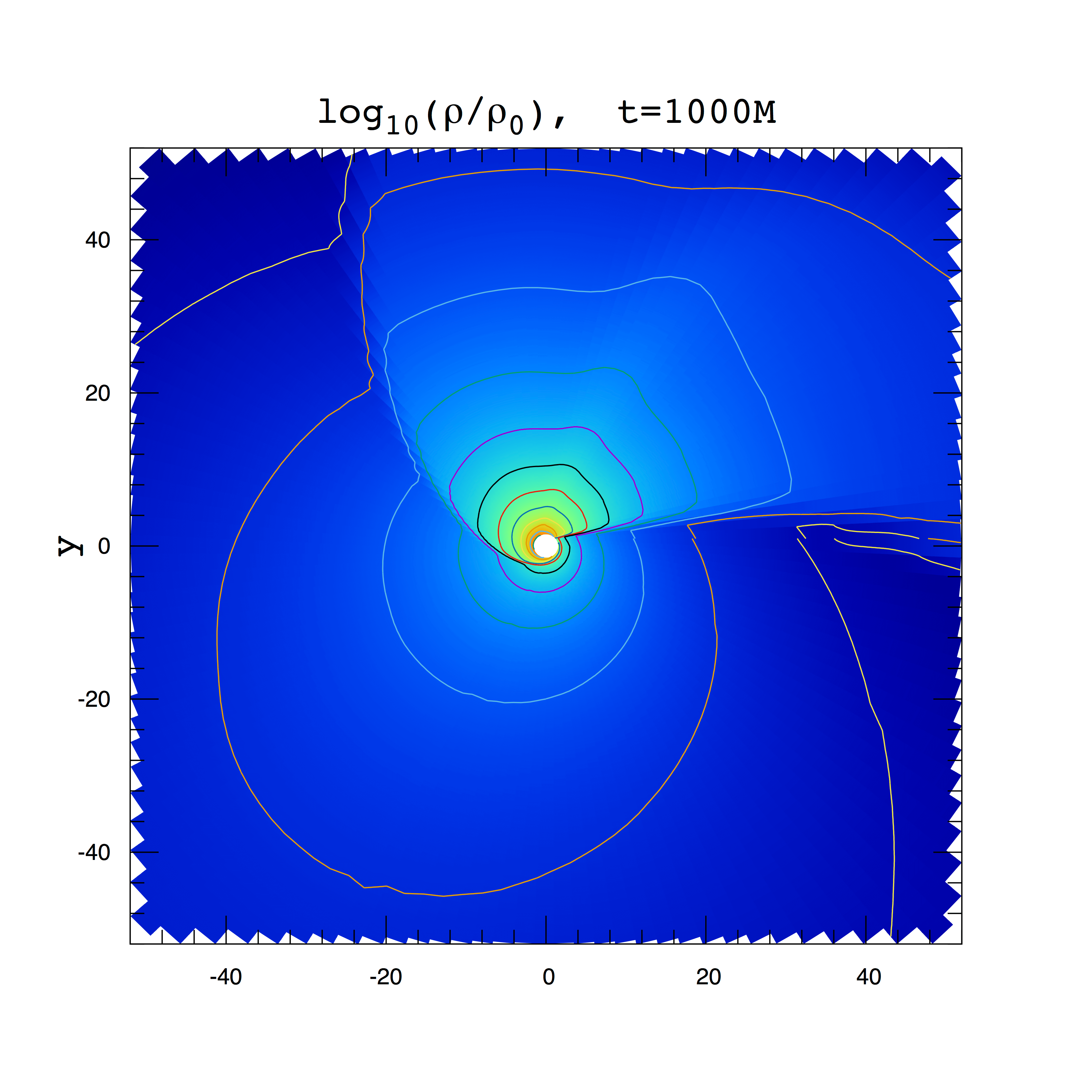} \hspace{-1.25cm}
\includegraphics[width=0.75\columnwidth]{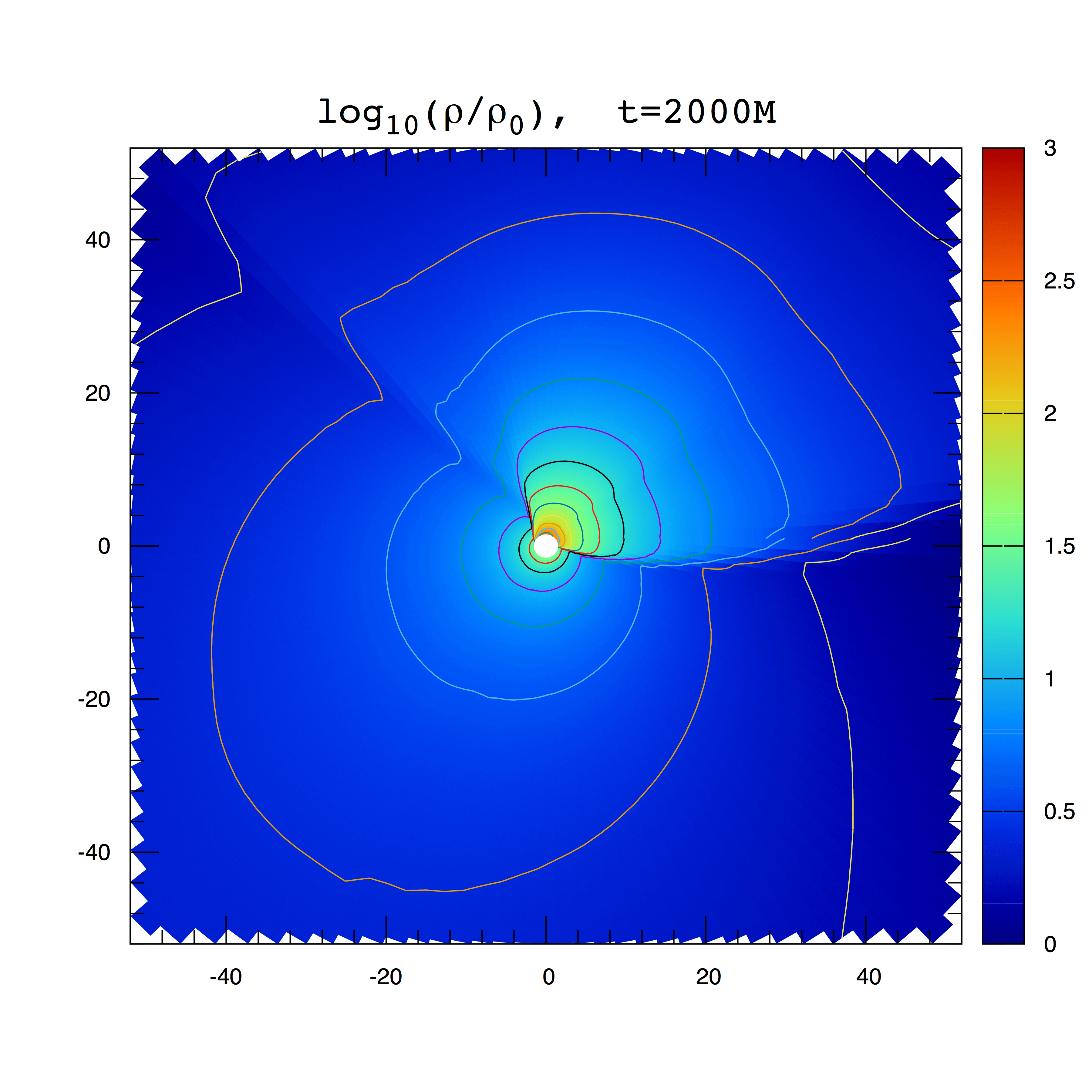}\\ 
\vspace{-0.5cm}
\includegraphics[width=0.75\columnwidth]{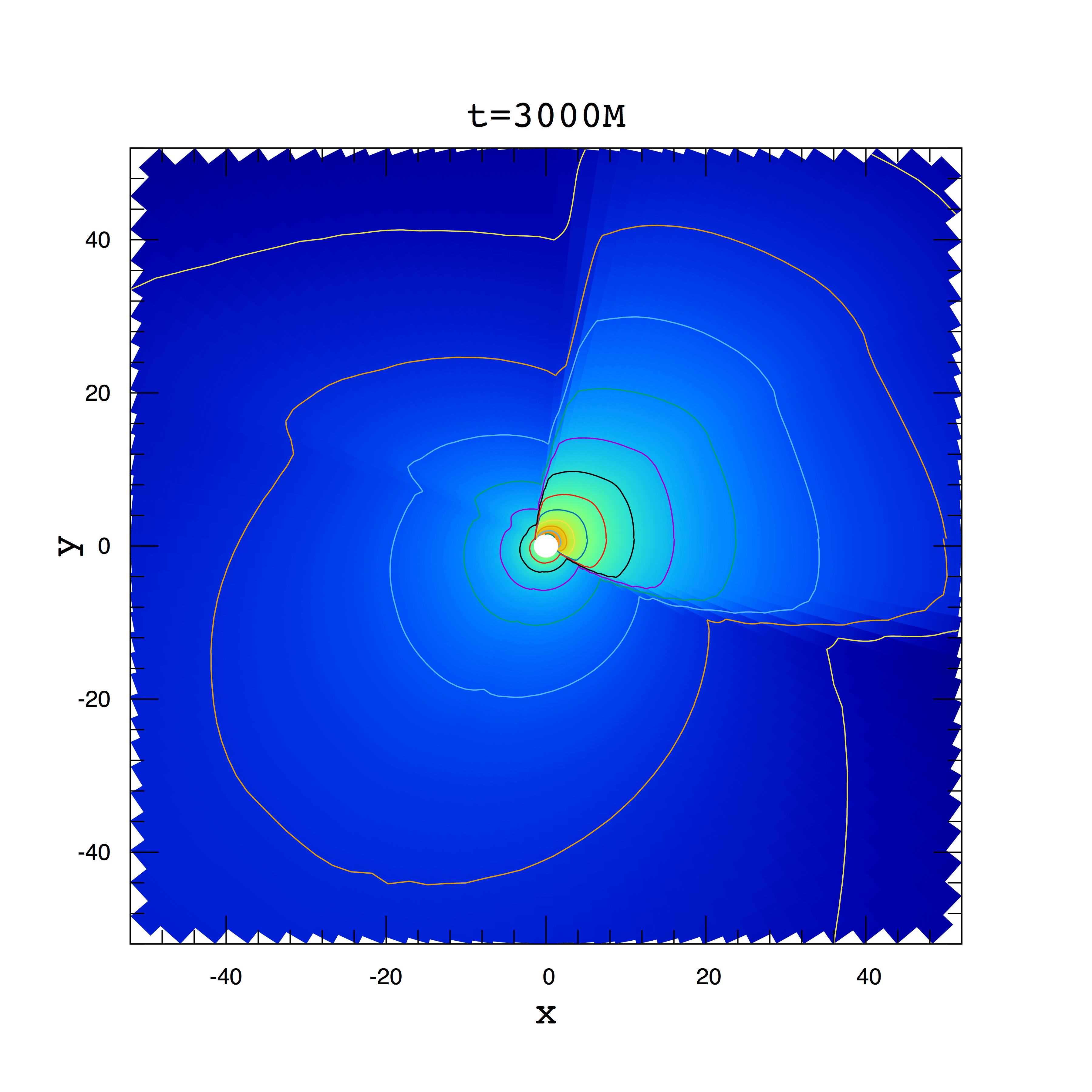}\hspace{-1.21cm} 
\includegraphics[width=0.75\columnwidth]{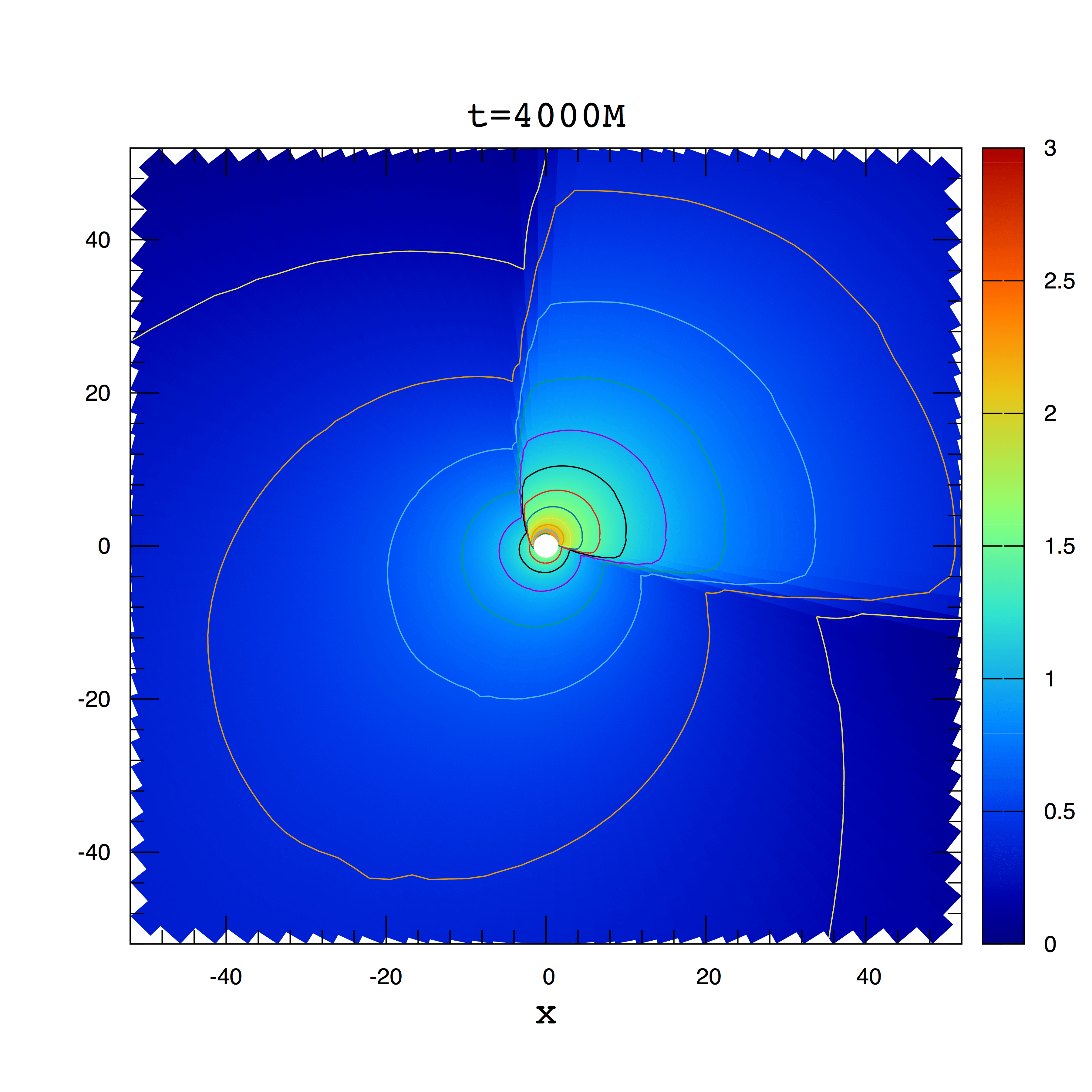} \\
%\vspace{-0.5cm}
%\includegraphics[width=0.55\columnwidth]{den_a098_erho05_ev03_t5.png}\hspace{-0.68cm} 
%{\vspace{-0.9cm}\includegraphics[width=0.45\columnwidth]{den_a098_erho05_ev03_t6.png}}
%\vspace{0.75cm}
\caption{Morphology of the rest mass density at different times, showing the accretion stages of a wind with Mach number ${\cal M}_{\infty} = 1$, velocity gradient $\epsilon_{v}=0.3$ and density gradient parameter $\epsilon_{\rho}=0.5$. The spin of the black hole is $a=0.98$. The system presents, at the early stages, a type of flip-flop behavior in the shock cone. However, after $t=4000M$ the accretion reaches a fairly steady state.}
    \label{fig:3}
\end{figure*}

 We also carry out a numerical simulation in which we consider an extremely rotating black hole with spin $a=0.98$.  This special case includes velocity and density gradients and assumes that the gas is moving with low Mach number, ${\cal M}_{\infty}=1$. We take the greatest values of the gradient parameters we can achieve, that is $\epsilon_{v}=0.3$ and $\epsilon_{\rho}=0.5$ . The morphology of the gas moving toward a black hole can be found in Figure \ref{fig:3}, in which we can observe a kind of flip-flop oscillation of the shock cone that appears in the early satges of evolution until $t=4000M$; after that, the system reaches a stationary pattern.  
  
\begin{figure}
\includegraphics[width=\columnwidth]{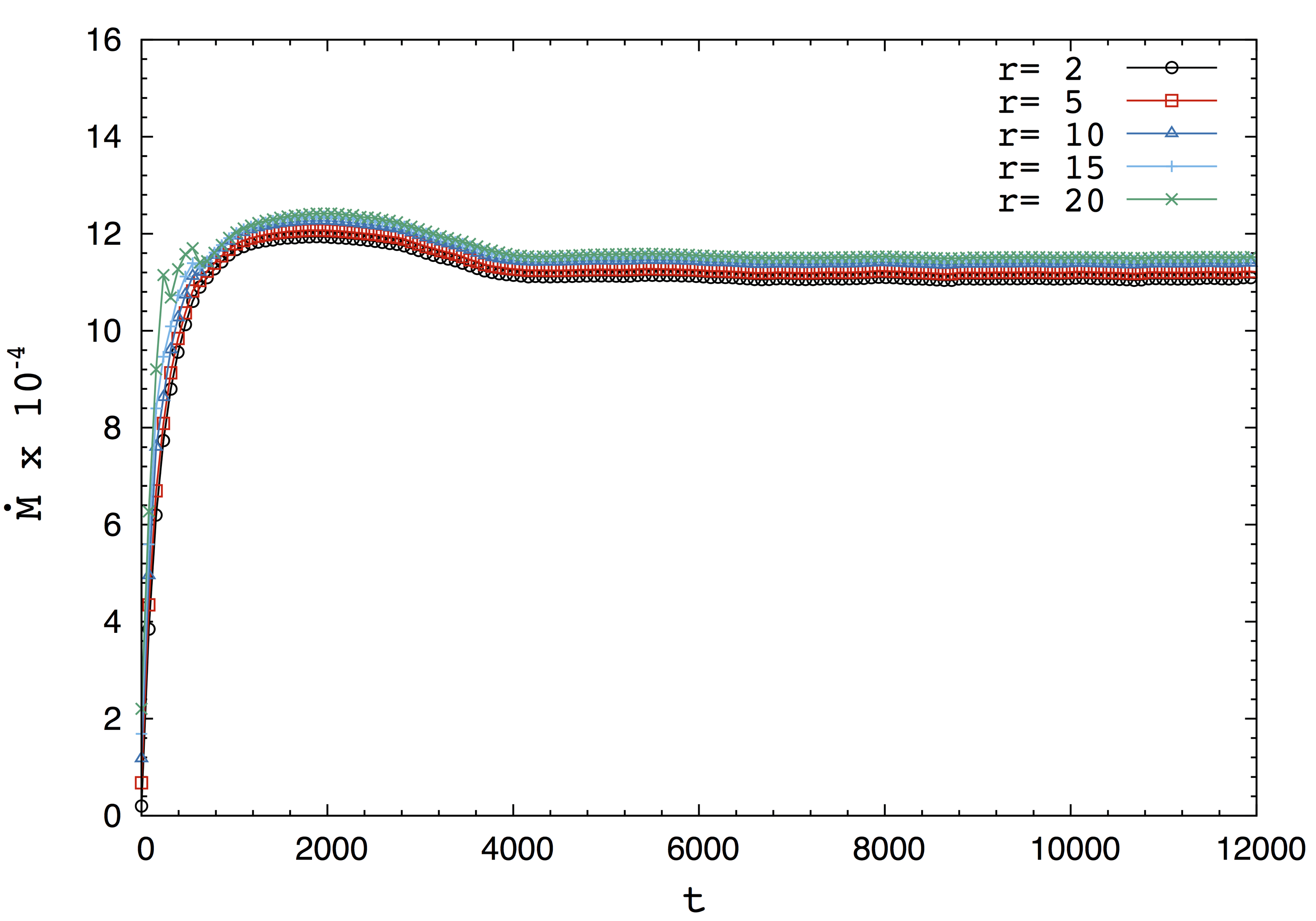} \\ 
\includegraphics[width=\columnwidth]{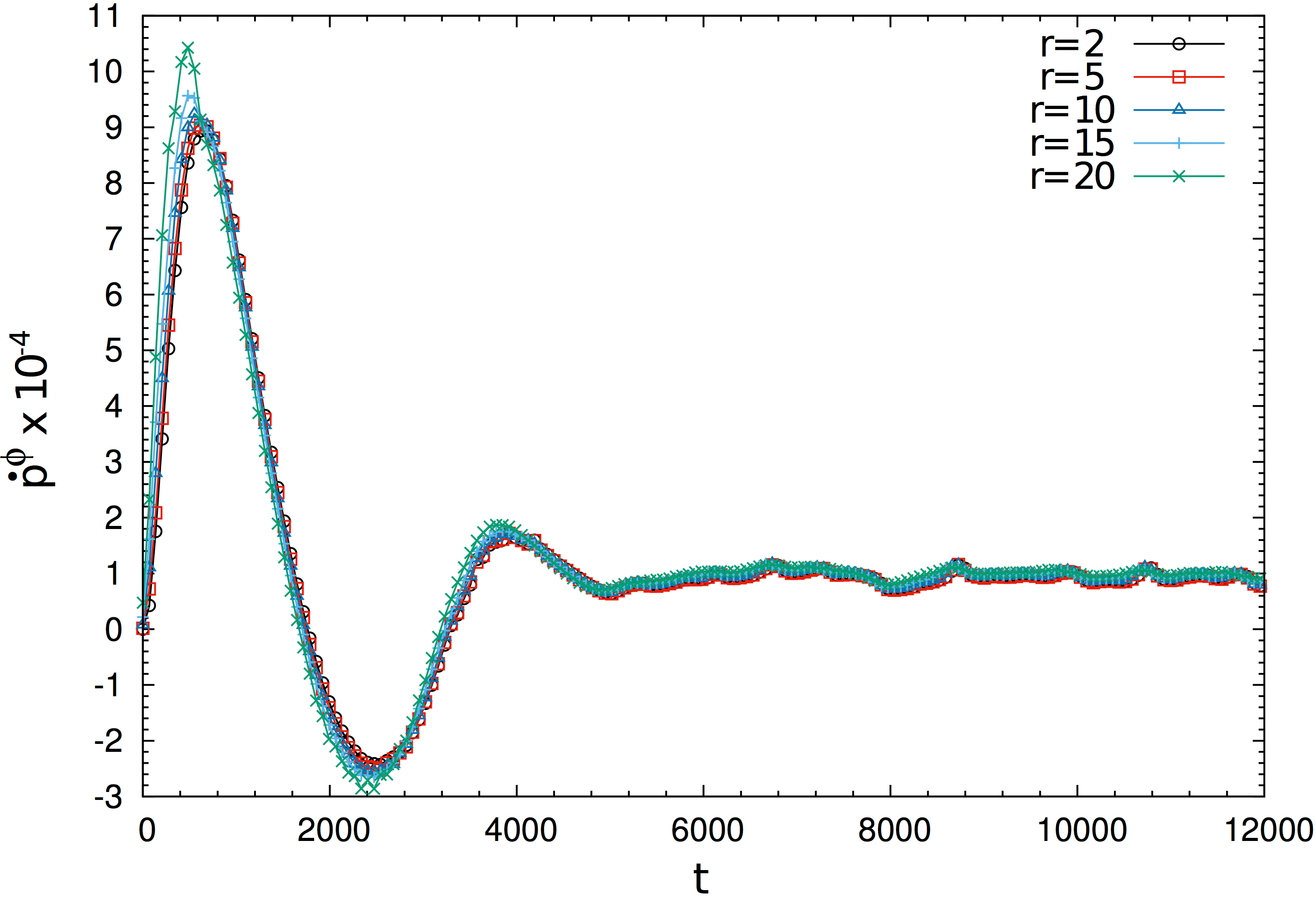}
    \caption{We show a long time evolution of the mass (top) and angular momentum (bottom) accretion rates for the model $M_7$. These quantities are measured at different detectors distributed uniformly along the radial coordinate, starting  near to the event horizon. In this case we assume that wind has not only velocity gradients but also density gradients. We adopted the particularly values $\epsilon_{v}=0.3$  and $\epsilon_{\rho}=0.5$ and ${\cal M}_{\infty} = 1$. }
    \label{fig:6}
\end{figure}

In the top panel of Figure \ref{fig:6}, we present the mass accretion rate measured by different detectors located in whole the domain. As we can see, $\dot{M}$ behaves similar in the different detectors, showing a stationary pattern.  On the other hand,  the most notable trends of unstable oscillations in angular momentum rates  (bottom panel) are found in Model $M_7$, in which we combine the density and velocity gradients with a high spin of the black hole, $a=0.98$. The long time evolution of this case can be observed in the bottom panel of Figure \ref{fig:6}, in which strong oscillations appear, ringing like a flip-flop that coincides with the oscillation in the morphology of the shock cone presented in Figure \ref{fig:3}. Nevertheless, after a few periods, the accretion rate reaches  a fairly steady state, fluctuating around the constant value, $\dot{p}^{\phi} \sim 1 \times 10^{-4}$.

\section{Conclusions}
\label{sec:conclus}

We have investigated, for the first time, the influence of the velocity gradients, with different  Mach numbers and adiabatic indexes, on the relativistic accretion of a supersonic gas moving onto a Schwarzschild and Kerr black holes with different spin parameters. The numerical evolutions were carried out on the equatorial plane using CAFE code. We particularly study velocity gradients, which are perpendicular to the relative motion of the relativistic fluid, similar to the Newtonian case \citep{1995A&A...295..108R,1997A&A...317..793R,1999A&A...346..861R}. Our numerical set of parameters focuses on the spin of the black hole $a$, velocity gradient $\epsilon_v$, particular Mach numbers, ${\cal M}_{\infty}=1,~3,~4,~5$ and adiabatic indexes ${\Gamma}=4/3,~5/3,~2$.

\begin{figure}
\includegraphics[width=\columnwidth]{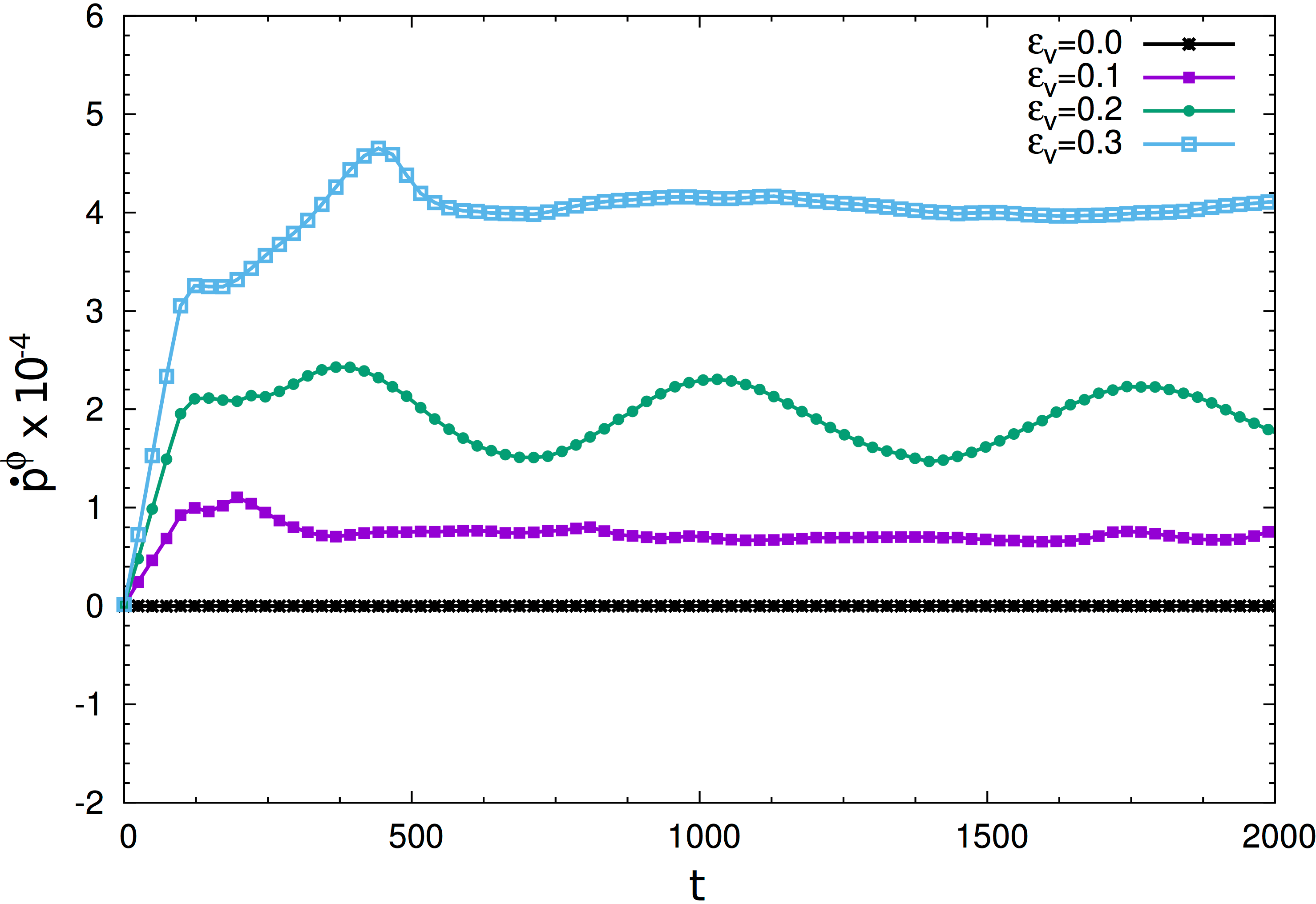} \\ 
\includegraphics[width=\columnwidth]{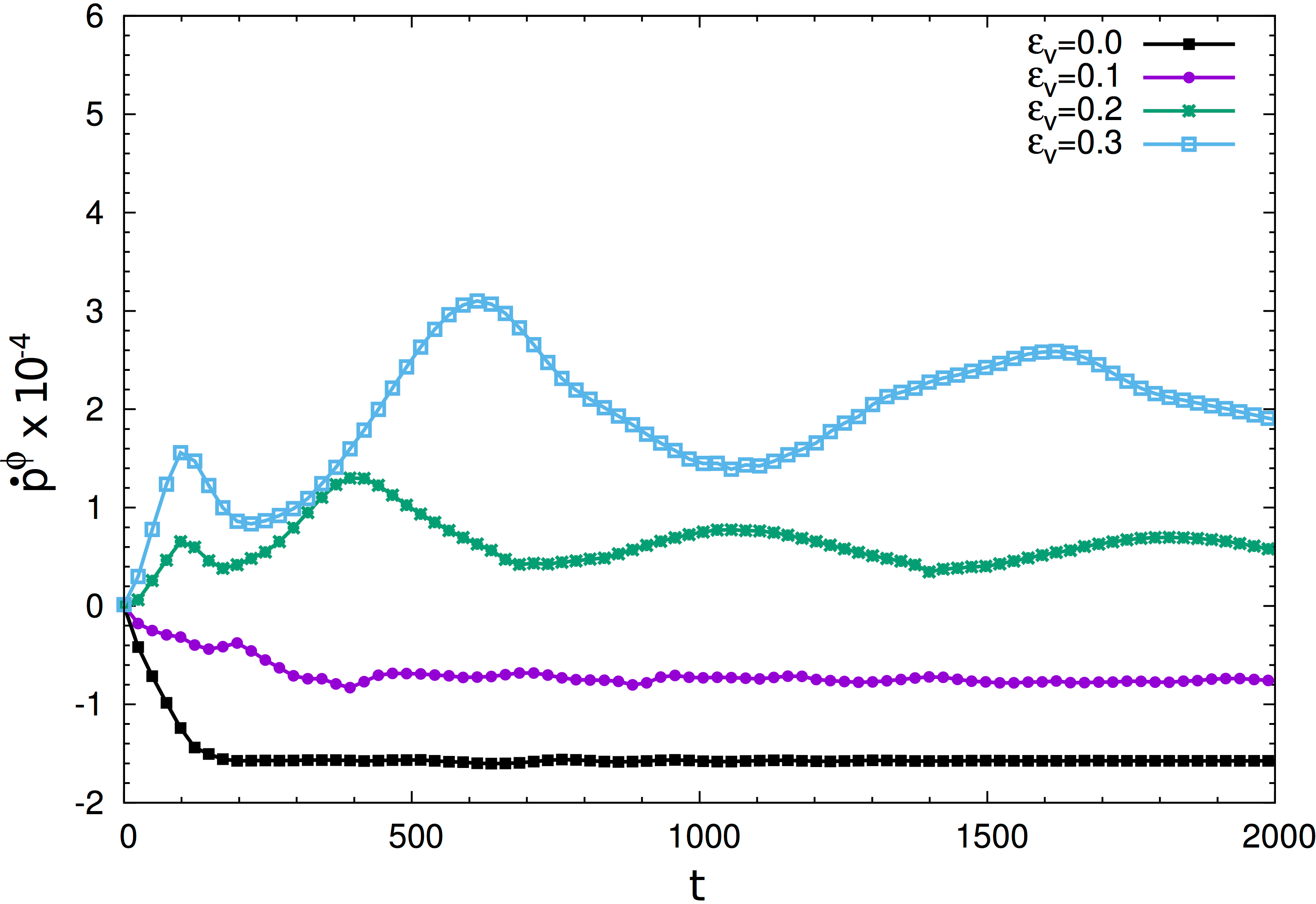}
    \caption{ This figure shows the angular momentum accretion rates measured near the event horizon. Again, as in the figure \ref{fig:4}, the top panel presents the case with spin zero, while in the bottom one with spin $a=0.5$.  We can appreciate notable changes when the velocity gradient parameter is increased, that is, the greater the velocity gradient is the greater the angular accretion rate.}
    \label{fig:5}
\end{figure}

\begin{figure*}
\includegraphics[width=8.5cm,height=6.5cm]{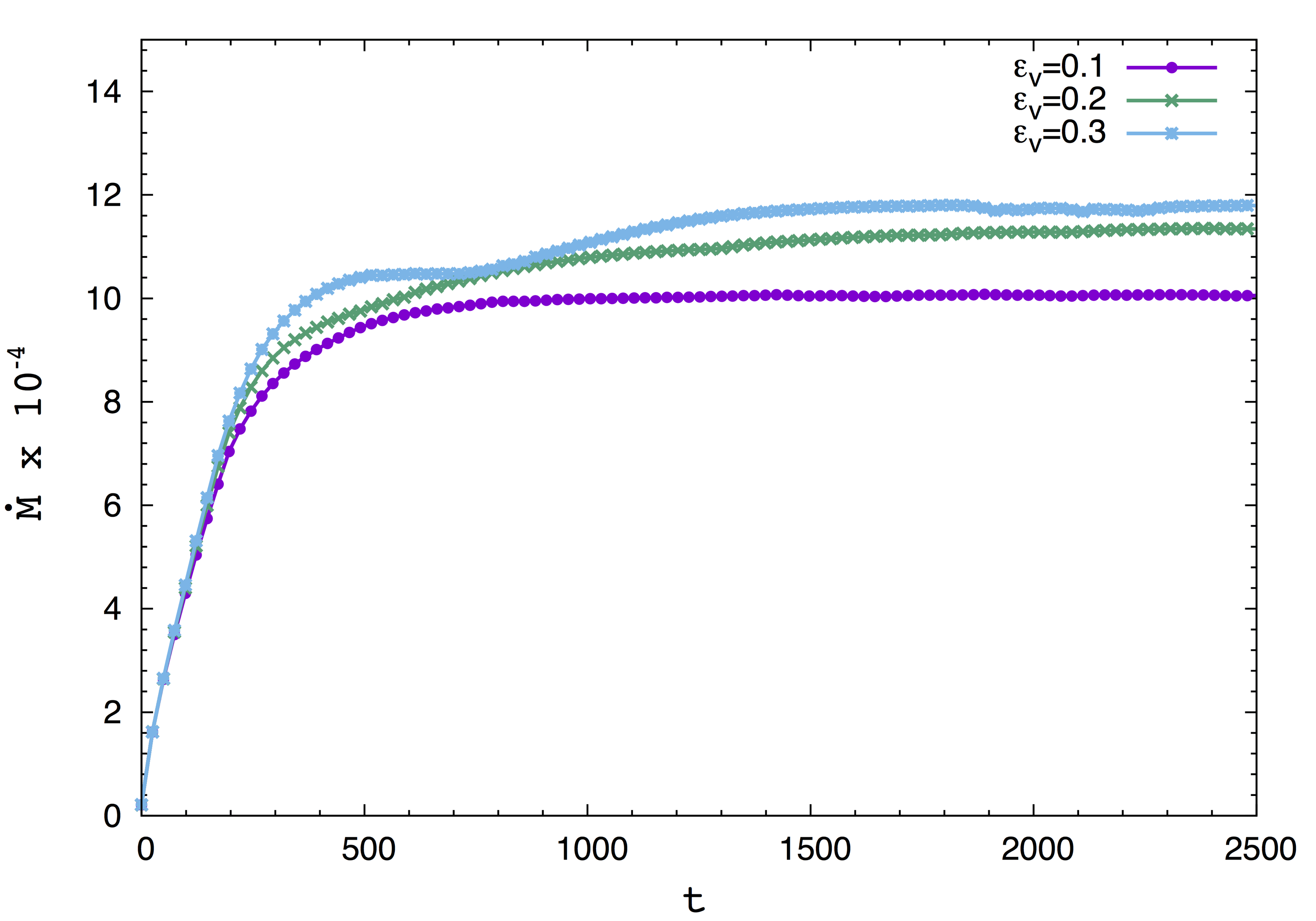} 
\includegraphics[width=8.5cm,height=6.5cm]{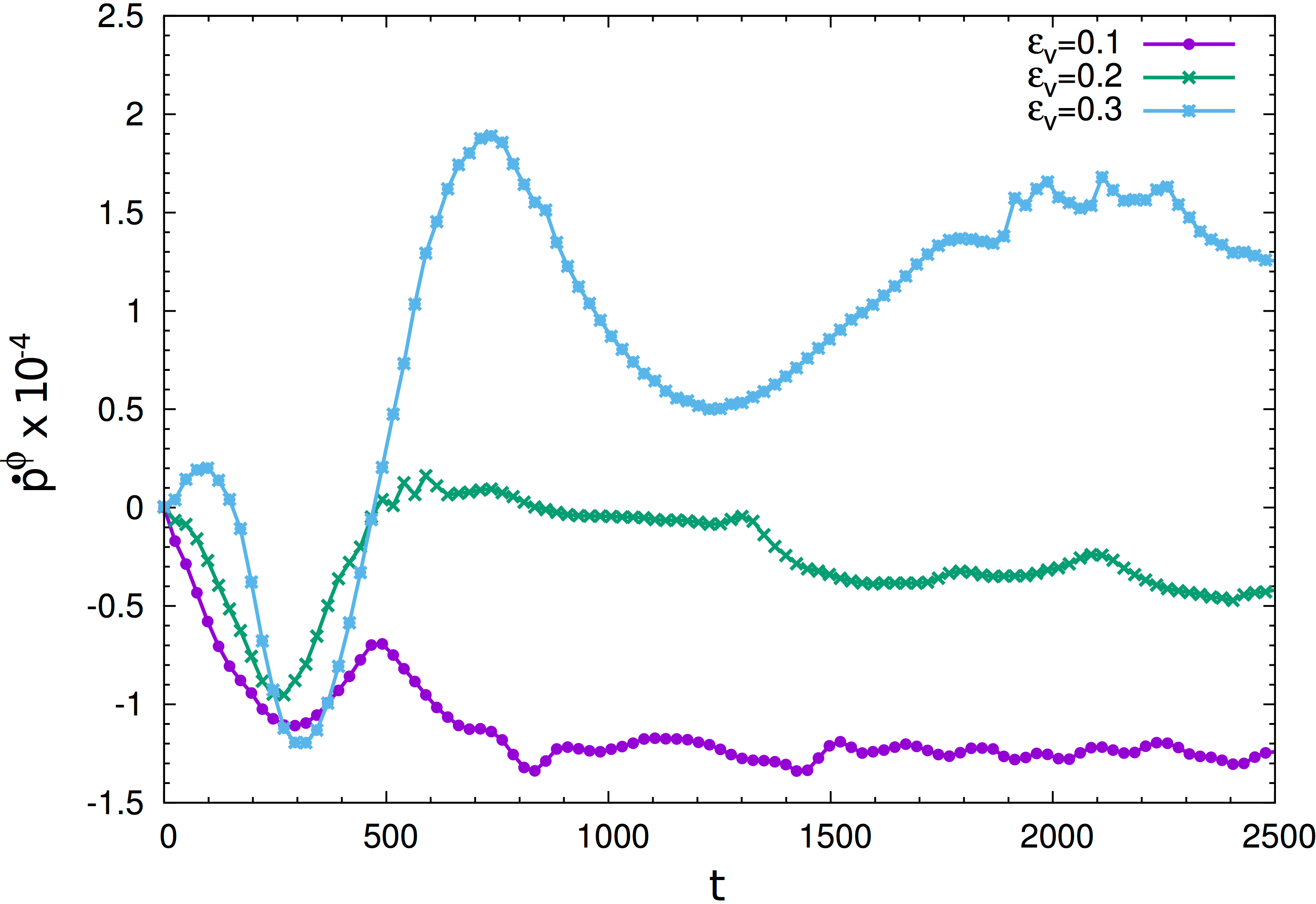}
\includegraphics[width=8.5cm,height=6.5cm]{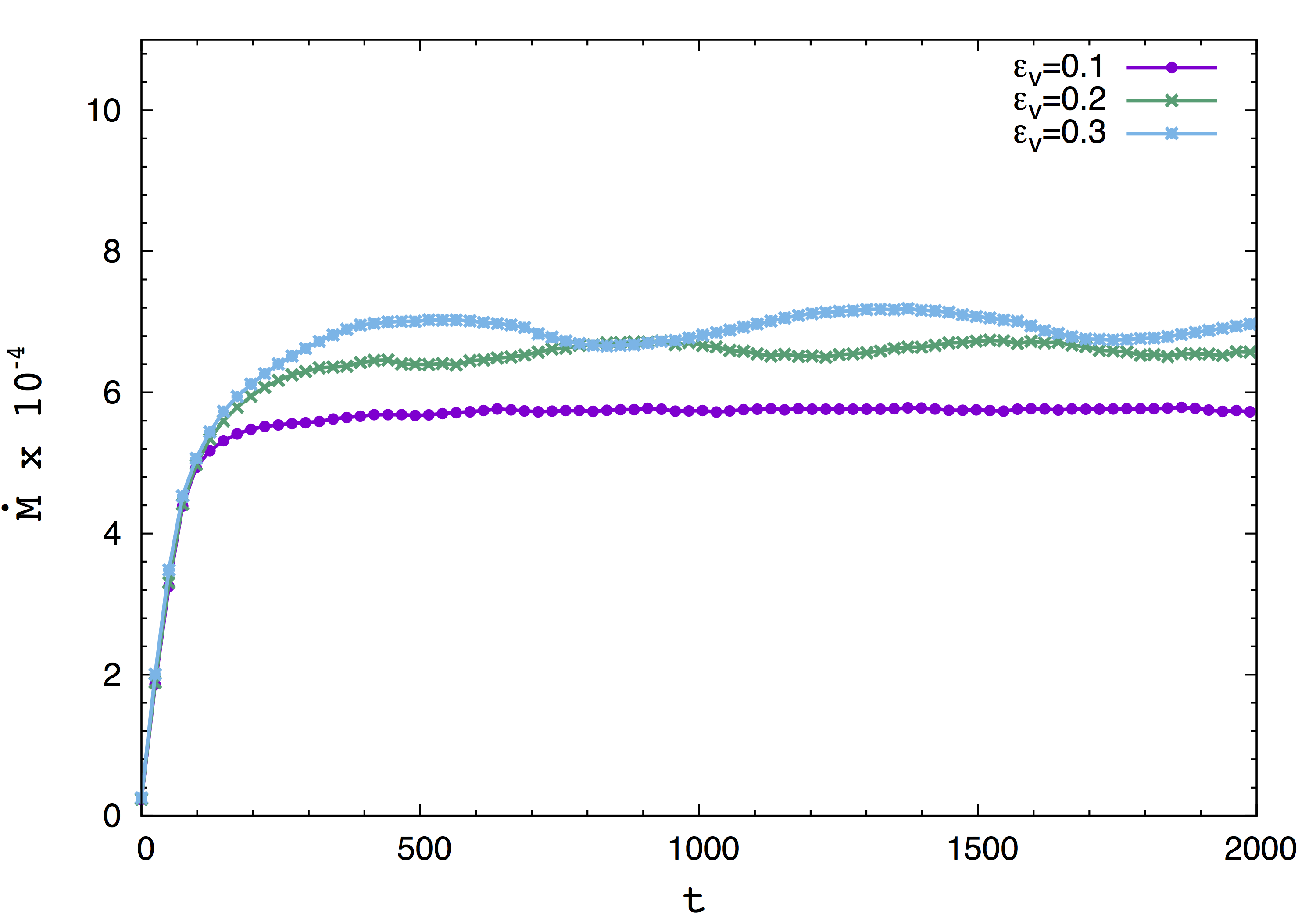} 
\includegraphics[width=8.5cm,height=6.5cm]{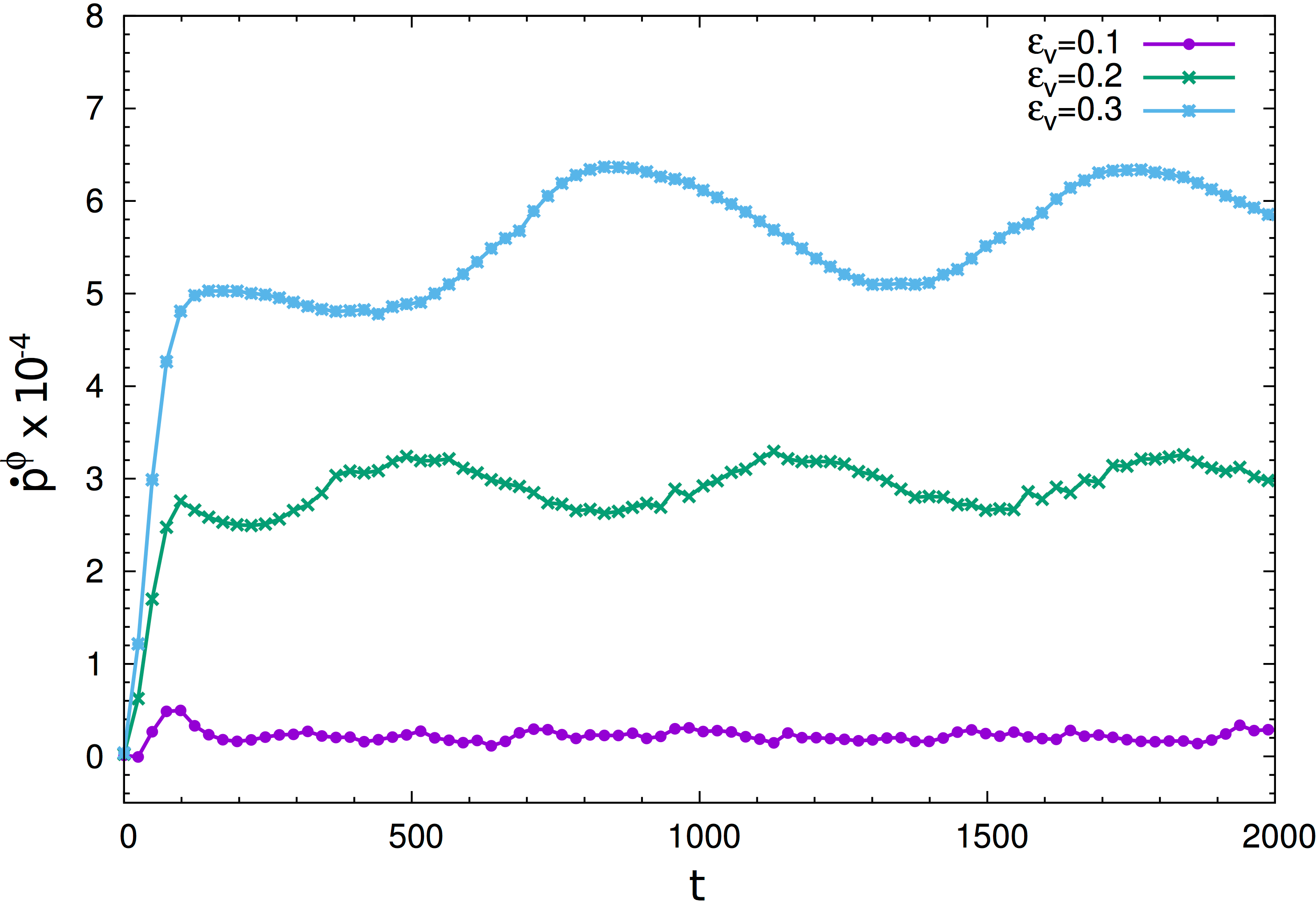}
    \caption{ In this figure, the mass (left column) and angular momentum (right column) accretion rates for a relativistic wind moving toward a rotating black hole ($a=0.5$) with asymptotical  Mach number ${\cal M}_{\infty} = 3$ (top row) and ${\cal M}_{\infty} = 5$ (bottom row), for different velocity gradient parameters, are presented. The behavior is similar to that presented in Figures \ref{fig:4} and \ref{fig:5}, that is,  the bigger the velocity gradient parameter is the bigger the mass and angular accretion rates are. On the other hand, despite the oscillations presented in the angular momentum accretion rates, the system in both cases reaches to a fairly stationary flow pattern.}
    \label{fig:7}
\end{figure*}

Our numerical results, for relativistic winds with Mach number ${\cal M}_{\infty}=3,4,5$,  are summarized in figures \ref{fig:2}, \ref{fig:4},  \ref{fig:5} and \ref{fig:7}.  These figures show the  morphology and the accretion rates for different values of the velocity gradient parameter $\epsilon_v=0,~0.1,~0.2,~0.3$, in combination with different spins of the black hole, $a=0,~0.5$. By comparing the morphology with and without velocity gradients, we find an evident dependence with the velocity gradients. Specifically, we obtain a notorious drag and greater opening angle on the resulting shock cone, when the gradients are considered. From figures  \ref{fig:4}, \ref{fig:5} and \ref{fig:7} and for the set of parameters used, we can observe that the greater the values of $\epsilon_v$ the greater  the mass and angular momentum accretion rates. Moreover, despite the oscillations presented in the angular accretion rates along the time evolution, the system reaches a fairly steady pattern. 

On the other hand, we have also studied a model where not only velocity gradients are considered but also density gradients. For this evolution, we consider a relativistic fluid, with Mach number ${\cal M}_{\infty}=1$, moving towards a black hole with a spin of $a=0.98$. The gradient parameters are $\epsilon_v=0.3$ and $\epsilon_{\rho}=0.5$. This specific system reveals, a kind of flip-flop behavior at the early stages of the evolution, see Figure \ref{fig:3}. However, after that the system reaches a stationary state.  This fact is more notorious in the angular momentum rate plot (Figure \ref{fig:6} (bottom)), where we can observe that in the early times until $t=4000M$ the system oscillates and after that, it relaxes reaching a steady state. 

Even thought, this work was carried out in the slab symmetry, we provide a novel vestige about the behavior of the accretion process in the presence of velocity and density gradients. In order to obtain more realistic astrophysical sceneries, it is necessary to consider a non-fixed background in full nonlinear general relativity, where the gravitational wave signature can be studied.  

\section*{Acknowledgements}

A. C-O gratefully acknowledges DGAPA postdoctoral grant to UNAM and financial support from CONACyT 57584 and CONACyT project 165584. F.D.L-C gratefully acknowledges the financial support from Colciencias (Programa Es Tiempo de Volver) and Vicerrector\'ia de Investigaci\'on y Extensi\'on,  Universidad Industrial de Santander, grant number 1822.  The numerical simulations are carried out on TYCHO cluster at IA-UNAM.

\bibliographystyle{mnras}
%\bibliography{bibliography} 


\bsp	\label{lastpage}
\end{document}